\documentclass[12pt, draftclsnofoot, onecolumn]{IEEEtran}
\pdfoutput=1
\usepackage[noadjust]{cite}
\usepackage{amsmath}
\usepackage{breqn}
\usepackage{array}
\usepackage{mdwmath}
\usepackage{amssymb}
\usepackage{algorithm}
\usepackage[noend]{algpseudocode}
\usepackage{eqparbox}
\usepackage{stfloats}
\usepackage{url}
\usepackage{tikz}
\usepackage{pgfplots}
\usepackage{float}
\graphicspath{{../pdf/}{../jpeg/}}
\DeclareGraphicsExtensions{.pdf,.jpeg,.png}
\usepackage{epstopdf}
\usepackage{amsfonts}
\usepackage{graphicx}
\usepackage[caption=false]{subfig}

\usepackage{pgfplots}
\usepackage{caption}
\usepackage{tikz,pgfplots}
\pgfplotsset{compat=1.11}

\newtheorem{lemma}{Lemma}
\newtheorem{theorem}{Theorem}

\newtheorem{corollary}{Corollary}

\begin{document}

\title{Distributed Reconfigurable Intelligent Surfaces Assisted Wireless Communication: Asymptotic Analysis under Imperfect CSI}
\author{Bayan Al-Nahhas,~\IEEEmembership{Student Member,~IEEE,} Qurrat-Ul-Ain~Nadeem,~\IEEEmembership{Member,~IEEE,} and Anas~Chaaban,~\IEEEmembership{Senior Member,~IEEE}
\thanks{The authors are with School of Engineering, The University of British Columbia, Kelowna, Canada.  (e-mail: bayanalnahhas@gmail.com, \{qurrat.nadeem, anas.chaaban\}@ubc.ca). Part of this paper has been presented in \cite{bay}.}
}

\maketitle
\vspace{-.2in}
\begin{abstract}
\vspace{-.05in}
This work studies the net sum-rate performance of a distributed reconfigurable intelligent surfaces (RISs)-assisted multi-user multiple-input-single-output (MISO) downlink communication system under imperfect instantaneous-channel state information (I-CSI) to implement precoding at the base station (BS) and statistical-CSI (S-CSI) to design the RISs phase-shifts. Two channel estimation (CE) protocols are considered for I-CSI acquisition: (i) a full CE protocol that estimates all direct and RISs-assisted channels over multiple training sub-phases, and (ii) a low-overhead direct estimation (DE) protocol that estimates the end-to-end channel in a single sub-phase. We derive the deterministic equivalents of    signal-to-interference-plus-noise ratio (SINR) and  ergodic net sum-rate under  Rayleigh and Rician fading and both CE protocols, for given RISs phase-shifts, which are then optimized based on S-CSI. Simulation results reveal that the low-complexity DE protocol yields better net sum-rate  than the full CE protocol when used to obtain CSI for precoding. A benchmark full I-CSI based RISs design is also outlined and shown to yield higher SINR  but lower net sum-rate than the S-CSI based RISs design due to the large overhead associated with full I-CSI acquisition. Therefore the proposed DE-S-CSI based design for precoding and reflect beamforming achieves high net sum-rate with low complexity, overhead and power consumption. 

\end{abstract}

\vspace{-.1in}
\begin{IEEEkeywords}
Reconfigurable intelligent surface (RIS), channel estimation (CE), instantaneous channel state information (I-CSI), statistical CSI (S-CSI), Rician fading, optimization.
\end{IEEEkeywords}

\section{Introduction}

In order to meet the requirements of high data rates and low  power consumption for the next generation  communication systems, an important challenge is to utilize energy-efficient technologies that can  provide dynamic control over the propagation of radio waves in different propagation environments. A transformative solution that addresses this challenge is to deploy reconfigurable intelligent surfaces (RISs) \cite{Tow_IRS, 6G2, 6G3,RIS_n1} on structures in the environment to  customize the propagation of radio waves through controlled reflections. Specifically, an RIS constitutes of a large number of  low-cost passive reflecting elements that  induce  phase shifts onto the incident signals, that can be  smartly tuned to realize desired communication objectives \cite{RIS_n1}. For example, the works in \cite{Guo, Wu1} and \cite{annie}  jointly optimize the precoding at the base station (BS) and passive reflect beamforming at the RIS to solve the sum-rate maximization problem, the  transmit power minimization problem and the minimum rate maximization problem respectively for the RIS-assisted multiple-input single-output (MISO) system. The authors in \cite{huang} proposed designs for   power allocation at the BS and  phases-shifts at the RIS elements to maximize the energy efficiency of the RIS-assisted multi-user MISO system that employs zero-forcing precoding.



While significant performance gains have been shown using a single RIS, the use of  distributed RISs in wireless communication settings  can  more effectively enhance coverage,  enable communication when multiple direct BS-users links are weak or blocked, improve the rank of the overall channel, and increase the spectral and energy efficiency of the system \cite{IRSstat, sumrate_dis}.  Some notable works studying distributed RISs-assisted communication systems include \cite{IRSstat,SWIPT_dis,sumrate_dis,maxmin_dis2, maxmin_dis, maxmin_dis3}. The authors in \cite{SWIPT_dis} studied a multiple RISs-assisted simultaneous wireless information and power transfer (SWIPT) system and proposed joint active and passive beamforming designs to minimize the transmit power at the BS subject to signal-to-interference-plus-noise ratio (SINR) constraints at information users and energy harvesting constraints at energy users. The authors in \cite{sumrate_dis} consider multiple single-antenna source-destination pairs assisted by a distributed RISs network and maximize the sum-rate by optimizing transmit power at sources and phase shifts matrices at the RISs.  The authors in  \cite{maxmin_dis2} consider a wireless network where multiple BSs serve their associated single-antenna users with the aid of distributed RISs. Considering Rayleigh fading and maximum ratio transmission (MRT) precoding, they derive an average-signal-to-average-interference-plus-noise ratio (ASAINR) expression in terms of RIS-user association parameters, which are then optimized  to maximize the minimum ASAINR among all users. 



Most of the current literature on RIS-assisted  systems assumes perfect channel state information (CSI) of all links to be available for beamforming design, which is very impractical given the RIS elements have no radio resources to send, receive or process pilot symbols, rendering channel estimation (CE) very challenging. To address this limitation, CE algorithms that exploit the sparsity of the RIS-assisted channel have been proposed  in \cite{CE_MU, cas}. Further, least-square and minimum mean squared error (MMSE) channel estimates of the direct BS-user links and RIS-assisted links have been derived in \cite{LS, LS1} and \cite{annie_OJ}, using a  pilot training based CE protocol requiring $N+1$ training sub-phases in each coherence block, where $N$ is the number of RIS elements.     There are some works that develop lower overhead CE protocols, for example: the idea of grouping adjacent RIS elements into sub-surfaces was introduced in \cite{ofdm_irs} and \cite{OFDM}, which decreases the training overhead but also reduces the  beamforming gains.  More recently, the authors in \cite{CER1} and \cite{CER2} propose three-phase and two-phase CE frameworks respectively,  in which the RIS-assisted channels of a typical user are estimated in the first phase, while the channels of other users are estimated with lower overhead in the next phase. The authors in \cite{hiba_CE} propose an MMSE-discrete Fourier transform (DFT) based CE protocol that exploits the static nature of the BS-RIS channel to reduce the CE overhead.  However, these protocols still require the training time to grow proportionally large with $N$, which  compromises the   net sum-rate.


Moreover, even if we overlook the large training overhead associated with estimating all direct and RIS-assisted channels and assume instantaneous CE, optimizing the RIS phase-shifts based on instantaneous CSI (I-CSI) at the pace of a fast-fading channel significantly increases the system complexity. To mitigate these challenges, the authors in \cite{IRSstat,IRSstat1, IRSstat2, IRSstat3,maxmin_dis3} design the RIS parameters using only statistical CSI (S-CSI) without requiring the I-CSI of individual BS-RIS and RIS-users channels. The only I-CSI then needed is of the aggregate end-to-end channel to design beamforming at the BSs. Since the S-CSI changes at a much slower pace than the I-CSI,  the RIS designs based on S-CSI not only reduce the training overhead associated with  estimation of all links, but  also  relax  the need for frequently reconfiguring the RISs. In this context, the authors in \cite{IRSstat} study a multi-RIS assisted multi-user MIMO system and optimize  the beamforming vectors at the BS and users as well as the RISs to maximize the sum-rate, without requiring  I-CSI of all involved channels. The proposed approach is implemented in an offline phase in which RIS phases are optimized relying only on statistical distribution of the locations of the users, and an online phase in which the beamforming at BS and users is optimized based on I-CSI of the end-to-end channel. The authors in  \cite{maxmin_dis3}   consider a two-timescale beamforming scheme for a distributed RISs-assisted multi-user MISO communication system that relies on the knowledge of second-order channel statistics  to design the RIS phase shifts and I-CSI of the aggregate end-to-end channel to implement MRT precoding. However the  phase shifts at each RIS are optimized to maximize the received energy at the closest user and not the sum-rate, and inter-user interference is tackled by controlling the positions at which RISs are deployed. 


While the works in \cite{IRSstat} and \cite{maxmin_dis3} have studied  multi-user wireless communication systems under I-CSI based precoding implementation at the BS and S-CSI based RIS phase shifts design, they assumed the I-CSI of the aggregate end-to-end channel  to  be perfectly known for the  implementation of precoding, which is not a practical assumption. Secondly, they did not compare the performance of S-CSI and  I-CSI based RIS designs in terms of training overhead, SINR and net sum-rate performance to draw insights as to which scheme performs better in different  operating regimes. These constitute the main questions of this paper, in which we study  a distributed RISs-assisted  multi-user  MISO communication system under Rayleigh and Rician fading channel models,  considering the scenarios of (i) full imperfect I-CSI versus aggregate imperfect I-CSI availability at the BS to implement precoding, and (ii) I-CSI versus S-CSI availability to design the RIS phase-shifts. To this end, we  analytically study the ergodic net sum-rate performance achievable under MRT precoding,  that is implemented  at the BS using imperfect I-CSI of the end-to-end channel obtained using two different CE methods, and utilize the derived expressions to optimize the RIS phase shifts based on S-CSI. Later we also study the net-sum rate performance when RIS phase shifts are designed based on imperfect I-CSI. The results help us investigate whether  the net sum-rate  is better when we acquire I-CSI of  the aggregate  end-to-end channel or when we acquire  I-CSI of all individual channels that constitute the end-to-end channel, while accounting for the training overheads of both methods. 



We consider two  CE protocols to obtain  I-CSI for precoding. The first acquires full I-CSI using the MMSE-DFT CE protocol  from \cite{hiba_CE} that  exploits the static nature of BS-RISs channels to estimate all  RISs-user channels and  direct BS-user channel, and uses them to construct the aggregate BS-user channel estimate. The estimates are computed under an optimized RIS phase shifts solution  that minimizes the CE error. While the  number of  training symbols required by this protocol is less as compared to others  \cite{annie_OJ, LS, LS1, CER1, CER2}, it still scales linearly with the number of RIS elements.  Recognizing this large overhead, we consider a second CE protocol, the direct estimation (DE) protocol, which estimates each  end-to-end BS-user channel in a single sub-phase for given RISs  phase shifts matrices, which have been optimized based on S-CSI to maximize the net sum-rate. The derived estimates under both protocols are used to implement MRT precoding at the BS. While  DE  is not useful for designing RISs phase shifts instantaneously, as  that requires knowledge of  all direct and RISs-assisted channels, we show that it is a viable protocol when combined with the S-CSI based RISs design, especially for large system sizes.

To do this, we resort to  asymptotic analysis and develop deterministic equivalents of the ergodic achievable net sum-rate under both MMSE-DFT and DE CE protocols for implementing MRT precoding for given RISs phase-shifts matrices.  Under Rician fading, the derived deterministic equivalents turn out to be   functions of the large-scale channel statistics as well as the RISs phase-shifts  that are then optimized  using a projected gradient ascent algorithm based on S-CSI. Under Rayleigh fading, the deterministic equivalents do not depend on the RISs phase-shifts, implying that the RISs do not yield  reflect beamforming gains under Rayleigh fading, but we show that they still yield an array gain which becomes significant in noise-limited systems.  As a performance benchmark, we also formulate an instantaneous achievable   net-sum rate maximization problem to design the RISs phase shifts based on the full I-CSI of all channels obtained using the MMSE-DFT CE protocol.   This work results in the following technical contributions:



\begin{itemize}
\item Explicit expressions of the MMSE estimates  of each BS-user, RIS-user as well as the resulting aggregate end-to-end channel  using the MMSE-DFT CE protocol under Rician  and Rayleigh fading, presented in Lemma 1 and Corollary 1 respectively. 

\item  Expressions of the MMSE estimates  of aggregate end-to-end channels  using the DE protocol under Rician and Rayleigh fading, presented in Lemma 3 and Corollary 3 respectively. 


\item Deterministic equivalents of the SINR and net sum-rate under Rician fading for given RISs phase shifts matrices. The results are presented for MMSE-DFT protocol  in Theorem 1, and for MMSE-DE protocol in Theorem 2.  The deterministic equivalents are simplified for Rayleigh fading  in Corollary 6 and 7 for MMSE-DFT and MMSE-DE protocols respectively. 

\item  A projected gradient ascent algorithm, outlined in Algorithm 1, to design the RISs phase shifts to maximize the ergodic net  sum-rate using the derived deterministic equivalents. The resulting algorithm only requires S-CSI for implementation which reduces complexity. 

\item Simulation results that  illustrate the excellent match yielded by the deterministic equivalents of the ergodic net sum-rate  for moderate system sizes, as well as the large gains that can be achieved by using passive low-power RISs. The results further reveal that:
\begin{enumerate}
\item The asymptotic net sum-rate is larger when using DE to acquire I-CSI for precoding as compared to when using MMSE-DFT protocol,  because the former  avoids the large overhead  associated with estimating all RISs-assisted channels. Since the net sum-rate  approaches a deterministic quantity for large systems, we can design the RISs phase shifts  using  S-CSI. As a result DE  suffices to obtain enough I-CSI to implement MRT.
\item The full I-CSI based RISs design that maximizes the instantaneous net sum-rate yields high SINR  but its net sum-rate performance is worse than the DE+S-CSI based scheme.
\item  Deploying  RISs yields both reflect beamforming and array gains under Rician fading channels that become dominant for higher Rician factors, while it yields only an array gain under Rayleigh fading  which becomes dominant in noise-limited scenarios.
\item Distributed RISs deployment outperforms centralized RIS deployment upto a certain distribution of RIS elements over multiple surfaces.
\end{enumerate}

\end{itemize}

The rest of the paper is organized as follows. Sec. II presents the system model and problem formulation. In Sec. III, we present the CE protocols and the asymptotic analysis of the net sum-rate under Rician and Rayleigh fading channels. The RISs  phase-shifts designs are proposed in Sec. IV. Simulation results and conclusions are provided in Sec. V and VI respectively.


\vspace{-.05in}
\section{System Model and Problem Formulation}

\begin{figure}[t!]
\centering
\includegraphics[scale=.28]{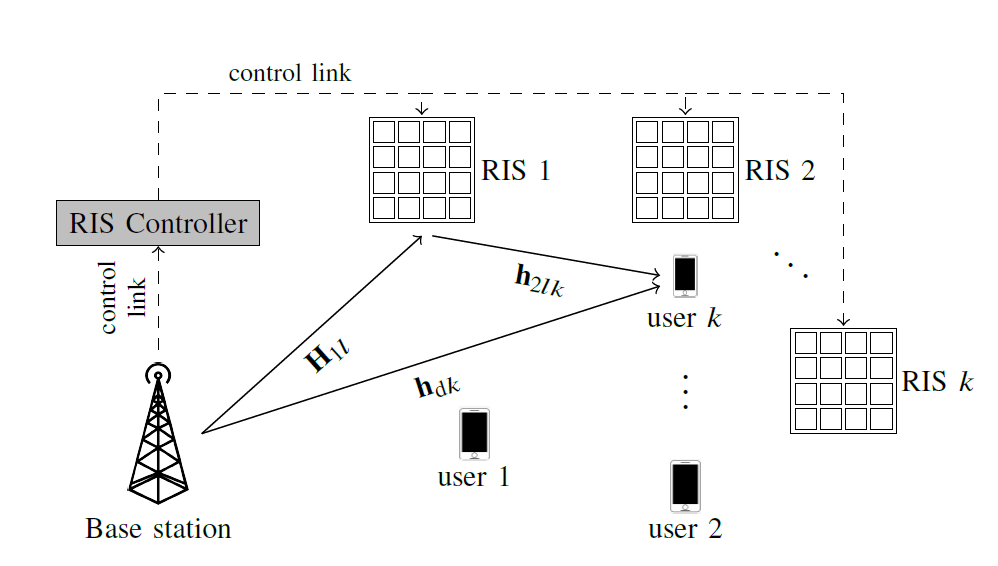}
\caption{Distributed RISs-assisted multi-user MISO system model. }
\label{LIS_model}
\end{figure}

As shown in Fig. \ref{LIS_model}, we consider a BS equipped with $M$ antennas communicating with $K$ single antenna users, with the assistance of  $L$ RISs composed of $N$  reflecting elements each that are connected to the BS via a control link. The signal model of this system is explained next. 

\vspace{-.05in}
\subsection{Signal Model}
\label{sysmodel}

 The BS wants to send information at rate $R_k$ to user $k$, $k=1,\dots,K$. To this end, the BS constructs codewords with symbols $s_k\sim \mathcal{CN}(0,1)$, and combines them into the transmit (Tx) signal vector $\mathbf{x}$ given as $\mathbf{x}=\sum_{k=1}^{K} \sqrt{p_{k}} \mathbf{g}_{k} s_{k}$, with $\mathbf{g}_{k} \in \mathbb{C}^{M\times 1}$ and $p_{k}>0$ being the precoding vector and signal power for user $k$ respectively. The Tx signal  satisfies the average Tx power constraint $ \mathbb{E}[||\mathbf{x}||^{2}]=\mathbb{E}[\text{tr}(\mathbf{P} \mathbf{G}^{H} \mathbf{G})] \leq  P_{max}$, where   $P_{max}$ is the Tx power budget, $\mathbf{P}=\text{diag}(p_{1}, \dots, p_{K})$ and $\mathbf{G}=[\mathbf{g}_{1}, \dots,\mathbf{g}_{K}]$. The received  signal $y_{k}$ at user $k$ is given as 
\begin{align}
\label{y_k}
&y_{k}= \mathbf{h}_k^{H} \mathbf{x}+n_k,
\end{align}
where $n_{k}\sim \mathcal{CN}(0,\sigma^2)$ is the noise, and  $\mathbf{h}_k$ is the channel between the BS and user $k$ given as
\begin{align}
\label{eq_ch}
&\mathbf{h}_{k}= \mathbf{h}_{dk}+\sum_{l=1}^L\mathbf{H}_{1l}\boldsymbol{\Theta}_l \mathbf{h}_{2lk},
\end{align} 
where $\mathbf{H}_{1l}\in \mathbb{C}^{M\times N}$ is the channel between the RIS $l$ and the BS, and $\mathbf{h}_{2lk} \in \mathbb{C}^{N\times 1}$ and  $\mathbf{h}_{dk}\in \mathbb{C}^{M\times 1}$  are the  channel vectors between user $k$ and  RIS $l$, and  user $k$ and the BS respectively. Also $\boldsymbol{\Theta}_l=\text{diag}(\phi_{l1},\ldots,\phi_{lN})$ represents the response of  RIS $l$, where $\phi_{ln}=\alpha_{ln} e^{j\theta_{ln}}$, $\theta_{ln}\in[0,2\pi] $ is the phase-shift introduced by  element $n$, and $\alpha_{ln}=1 $ is the amplitude reflection constant.

\subsection{Channel Models}


Each BS-RIS channel is considered to be line-of-sight (LoS) dominated. This assumption, made in many other related works \cite{Wu2, LoS1, LoS2, LoS4,  maxmin_dis, annie, annie_OJ}, is supported in the literature using two points. First, the LoS path between the BS and RIS can be guaranteed through  appropriate deployment of the RIS.   Second, the path loss for non-LoS (NLoS) paths is much larger than that for the LoS path in the next generation systems, with the typical value of  Rician factor reported as being between $20$\rm{dB} and $40$\rm{dB} \cite{LoS1}, which is sufficiently large to neglect any NLoS  paths in $\mathbf{H}_{1l}$. Under these remarks, we assume  BS-RIS channels to be LoS dominated as modeled next.

A uniform rectangular array (URA) of $N=N_1\times N_2$ reflecting elements is considered at each RIS, where $N_1$ and $N_2$ are the number of elements placed with inter-element spacing $d_{RIS}^{(1)}$ and $d_{RIS}^{(2)}$ along the two principal directions of the URA.  A uniform linear array (ULA) is considered at the BS, with  an antenna spacing of $d_{BS}$. Under the spherical wave model and considering a channel attenuation coefficient  of $\beta_{1l}$, the entries of the LoS channel $\mathbf{H}_{1l}$ are given as \cite{losura}
\begin{align}
\label{H_1}
&[\mathbf{H}_{1l}]_{m,n}=\sqrt{\beta_{1l}}\exp\left(j \frac{2\pi}{\lambda} \bar{l}_{(m),(n_1,n_2)_l}\right), 
\end{align}
 where $n=(n_1-1)N_2+n_2$, $n_1=1,\dots, N_1$, $n_2=1, \dots, N_2$,  and $\bar{l}_{(m),(n_1,n_2)_l}$ is the path length between  BS antenna $m$ and RIS $l$'s element $(n_1,n_2)$, given as $\bar{l}_{(m),(n_1,n_2)_l}=||\mathbf{a}_{RIS}^{(n_1,n_2)_l}-\mathbf{a}_{BS}^{(m)}||$. Here $\mathbf{a}_{RIS}^{(n_1,n_2)_l}$ is the steering vector from the global origin  to  RIS $l$'s element $(n_1, n_2)$, and $\mathbf{a}_{BS}^{(m)}$ is the steering vector from the global origin to  BS antenna $m$. The expressions of $\mathbf{a}_{BS}^{(m)}$ and $\mathbf{a}_{RIS}^{(n_1,n_2)_l}$,  found in \cite[equations (12) and (13)]{losura}, depend on $d_{RIS}^{(1)}$, $d_{RIS}^{(2)}$, $d_{BS}$, and the distance $\bar{D}_l$ between the BS and RIS $l$.  This model  is developed under the spherical wavefront assumption, and imposes no restriction on  rank of $\mathbf{H}_{1l}$, which can  be high for moderate $\bar{D}_l$ and large $N$ \cite{losura}. 

For the RIS-user channels $\mathbf{h}_{2lk}$ and the BS-user channels $\mathbf{h}_{dk}$,  we consider Rician and Rayleigh fading models. Under Rician fading, the channels are expressed as \vspace{-.1in}
\begin{align}
\label{ch111}
&\mathbf{h}_{2lk}^{\rm ric}= \mathbf{h}^{\rm n} _{2lk}+\bar{\mathbf{h}}_{2lk}, \hspace{.12in} \mathbf{h}_{dk}^{\rm ric}= \mathbf{h}^{\rm n}_{dk}+\bar{\mathbf{h}}_{dk}
\end{align}
where $\mathbf{h}_{2lk}^{\rm n}$ and $\mathbf{h}_{dk}^{\rm n}$ are the NLoS components, and $\bar{\mathbf{h}}_{2lk}$ and $\bar{\mathbf{h}}_{dk}$ are the LoS components. The NLoS components are given by 
 $\mathbf{h}^{\rm n}_{2lk}=\sqrt{\frac{1}{\kappa_{2lk}+1}}\mathbf{h}^{\rm ray}_{2lk}$ where $\mathbf{h}^{\rm ray}_{2lk} \sim \mathcal{CN}(0,\beta_{2lk} \mathbf{I}_N)$, and $\mathbf{h}^{\rm n}_{dk}=\sqrt{\frac{1}{\kappa_{dk}+1}}\mathbf{h}^{\rm ray}_{dk}$ where $\mathbf{h}^{\rm ray}_{dk}\sim\mathcal{CN}(0,\beta_{dk} \mathbf{I}_M)$. The LoS components are modeled as $\bar{\mathbf{h}}_{dk}=\sqrt{\frac{{\beta}_{dk}\kappa_{dk}}{\kappa_{dk}+1}}[1, e^{j2\pi d_{BS} \cos (\phi_{dk})},  \dots  , e^{j2\pi d_{BS} (M-1) \cos (\phi_{dk})}]$ and  $\bar{\mathbf{h}}_{2lk}=\sqrt{\frac{ {\beta}_{2lk} \kappa_{2lk}}{\kappa_{2lk}+1}} [\mathbf{b}_{2lkz} \otimes \mathbf{b}_{2lkx}]$, where $\mathbf{b}_{2lkz}=[1, e^{j2\pi d^{(1)}_{RIS} \cos (\phi_{2lk})},  \dots  , e^{j2\pi d_{RIS}^{(1)} (N_1-1) \cos (\phi_{2lk})}]$,  $\mathbf{b}_{2lkx}=[1, e^{j2\pi d^{(2)}_{RIS} \cos (\phi_{2lk})},  \dots  , \\ e^{j2\pi d_{RIS}^{(2)} (N_2-1) \cos (\phi_{2lk})}]$, and $\phi_{dk}$ and $\phi_{2lk}$ are the angles of departure (AoD) of the wave-vector from the BS and RIS $l$ to user $k$ respectively, defined in the azimuth plane from the x-axis. Moreover, $\beta_{dk}$ and $\beta_{2lk}$ are the channel attenuation factors, and  $\kappa_{dk}$ and $\kappa_{2lk}$ are the Rician factors for the BS-user $k$ link and RIS $l$-user $k$ link respectively. Under Rician fading, the channels are distributed as $\mathbf{h}_{2lk}^{\rm ric}\sim \mathcal{CN}\left(\bar{\mathbf{h}}_{2lk},\frac{\beta_{2lk}}{\kappa_{2lk}+1} \mathbf{I}_N\right)$, and $\mathbf{h}_{dk}^{\rm ric}\sim \mathcal{CN}\left(\bar{\mathbf{h}}_{dk},\frac{\beta_{dk}}{\kappa_{dk}+1} \mathbf{I}_M\right)$. The overall channel between user $k$ and the BS can be written as 
\begin{align}
\label{eq_ch1_RIC}
\mathbf{h}^{\rm ric}_{k}=\mathbf{h}^{\rm n}_{dk}+\bar{\mathbf{h}}_{dk}+\sum_{l=1}^L \mathbf{H}_{1l}\boldsymbol{\Theta}_l \mathbf{h}^{\rm n}_{2lk}+\sum_{l=1}^L \mathbf{H}_{1l}\boldsymbol{\Theta}_l \bar{\mathbf{h}}_{2lk}
\end{align} 
which is statistically equivalent to the following representation:  \vspace{-.1in}
\begin{align}
\label{eq_ch1_RIC22}
&\mathbf{h}^{\rm ric}_{k}=\bar{\mathbf{h}}_{dk}+\sum_{l=1}^L \mathbf{H}_{1l}\boldsymbol{\Theta}_l \bar{\mathbf{h}}_{2lk}+\mathbf{A}^{{\rm ric}^{1/2}}_{k}\mathbf{z}_{k},
\end{align}
where $\mathbf{z}_{k}\sim \mathcal{CN}(\mathbf{0}, \mathbf{I}_{M})$ and $\mathbf{A}^{\rm ric}_{k}=\frac{ \beta_{dk}}{\kappa_{dk}+1}\mathbf{I}_{M}+\sum_{l=1}^L \frac{\beta_{2lk}}{\kappa_{2lk}+1}\mathbf{H}_{1l} \mathbf{H}_{1l}^H$. 



The Rayleigh fading channels $\mathbf{h}^{\rm ray}_{dk}$ and $\mathbf{h}^{\rm ray}_{2lk}$ are as defined below \eqref{ch111}, and can also be obtained from the Rician channel models by  setting the Rician factors $\kappa_{dk}$ and $\kappa_{2lk}$ as $0$. 

\vspace{-.15in}
\subsection{Channel State Information and Precoding}
\vspace{-.05in}
There are two ways to estimate the channels in an RISs-assisted system. The first is to acquire the CSI of all individual RISs-assisted channels, i.e. the RIS-user channels $\mathbf{h}_{2lk}$'s and the direct BS-user  channels $\mathbf{h}_{dk}$'s, and use this full I-CSI for precoding at the BS as well as for optimizing the RISs phase shifts  to obtain favourable instantaneous channels. However, as discussed in the introduction, many existing CE protocols require a training time that grows proportionally with $N$  to acquire this CSI \cite{LS, LS1, annie_OJ, CER1, CER2, hiba_CE}. The second way is to acquire the I-CSI of only the aggregate end-to-end channel $\mathbf{h}_{k}$ for given  $\boldsymbol{\Theta}_l$'s that are optimized based on S-CSI to obtain favourable channel statistics. The BS then uses the estimated end-to-end $\mathbf{h}_{k}$  to implement precoding.  In this work we focus on the latter  and consider the first method as a benchmark. To implement precoding, we consider the BS to have imperfect I-CSI of each aggregate BS-user channel $\mathbf{h}_k$, $k=1,\dots, K$, represented as  $\mathbf{h}_k=\hat{\mathbf{h}}_{k}+\tilde{\mathbf{h}}_{k}$, where  $\hat{\mathbf{h}}_{k}$ is the channel estimate and $ \tilde{\mathbf{h}}_{k}$ is the  estimation error. Two CE protocols to estimate $\mathbf{h}_k$ will be discussed in Sec. III-A.


 
 The  estimate $\hat{\mathbf{h}}_k$ is used to implement precoding at the BS.  Linear precoding schemes like MRT and zero-forcing (ZF) are asymptotically optimal for a MISO broadcast channel as $M$  grows large \cite{massive1, massiveMIMO}. In this work, we focus on the large $(M,K)$ regime given the massive connectivity supported by 5G networks and consider MRT precoding, since it reduces the computational complexity greatly as compared to ZF that involves the inversion of the Gram matrix of joint users' channel matrices,  which has a prohibitive computational complexity proportional to $K^2 M$. The precoding vectors are given as ${\mathbf{g}}_{k} = \zeta \hat {\mathbf{h}}_{k}$, where $\zeta $ satisfies the power constraint  $ \mathbb{E}[||\mathbf{x}||^{2}] \leq  P_{max}$ as $\zeta^2 = P_{max}/\Psi$, where $\Psi =\mathbb{E}\left[{\rm{tr}\left( {\mathbf{P}\hat {\mathbf{H}}}{ \hat {\mathbf{H}}^H} \right)}\right]$ and $\hat{\mathbf{H}}^H=[\hat {\mathbf{h}}_{1}, \hat {\mathbf{h}}_{2} \dots \hat {\mathbf{h}}_{K}]\in \mathbb{C}^{M\times K}$.\footnote{ The extension of the analysis in this work to other precoding schemes like ZF is possible using the methodology in \cite{wag33}. However the resulting expressions will be  complex and will yield no direct insights into the impact of the RIS on the sum-rate.  }

\vspace{-.1in}

\subsection{Achievable Rate and Problem Formulation}

For the considered system model, we present an ergodic achievable net-rate expression for each user, exploiting a technique from \cite{medard}, which is widely applied in works on large-scale MIMO systems \cite{HJY, ML2}. The technique exploits the channel hardening property of large-scale MIMO systems which states that as $M$ grows large, the effective channel $\mathbf{h}_{k}^{H} \mathbf{g}_{k}$ of user $k$ approaches its average value $\mathbb{E}[\mathbf{h}_{k}^{H} \mathbf{g}_{k}]$. Under this property, the authors of  \cite{medard} assume  the availability of only S-CSI (i.e. knowledge of $\mathbb{E}[\mathbf{h}_{k}^{H} \mathbf{g}_{k}]$) at the users to compute the SINR.  The main idea then is to decompose $y_{k}$ in (\ref{y_k}) using the definition of $\mathbf{x}$  as $y_k=\sqrt{p_{k}} \mathbb{E}[\mathbf{h}_{k}^{H} \mathbf{g}_{k}] s_{k} + \sqrt{p_{k}} (\mathbf{h}_{k}^{H} \mathbf{g}_{k}-\mathbb{E}[\mathbf{h}_{k}^{H} \mathbf{g}_{k}]) s_{k} +\sum_{f\neq k} \sqrt{p_{f}} \mathbf{h}_{k}^{H} \mathbf{g}_{f} s_{f} +{n}_{k}$ and assume that the average effective channel $\mathbb{E}[\mathbf{h}_{k}^{H} \mathbf{g}_{k}]$ is perfectly known at user $k$. Then by treating  interference and channel uncertainty as worst-case independent Gaussian noise, we conclude that user $k$ can achieve the ergodic net rate 
\begin{align}
\label{rate_11}
R_{k}=\left(1-\frac{S \tau_S}{\tau_C}  \right) \log_2(1+\gamma_{k}),
\end{align}
where $S$ is the number of CE sub-phases of length $\tau_S$ symbols, $\tau_C$ is the length of each coherence block, and $\gamma_k$ is the downlink SINR of user $k$. The expression of $\gamma_k$, under MRT precoding defined in Sec. II-C, is obtained using the decomposition of  $y_k$ given above (7)  as  
\begin{align}
\label{SINR_MRT}
& \gamma_{k}=\frac{p_{k} |\mathbb{E}[\mathbf{h}_{k}^{H} \hat{\mathbf{h}}_{k}]|^{2}}{p_{k} \mathbb{V}\text{ar}[\mathbf{h}_{k}^{H} \hat{\mathbf{h}}_{k}] + \sum_{f\neq k} p_{f} \mathbb{E} [|\mathbf{h}_{k}^{H} \hat{\mathbf{h}}_{f}|^{2}]+ \frac{\Psi}{\rho}},
\end{align}
where $\rho=\frac{P_{max}}{\sigma^2}$ and $\Psi =\mathbb{E}\left[{\rm{tr}\left( {\mathbf{P}\hat {\mathbf{H}}}{ \hat {\mathbf{H}}^H} \right)}\right]$. The ergodic achievable  net sum-rate is  given as \vspace{-.05in}
\begin{equation}
\label{R_sum_MC}
R_{sum}= \sum_{k=1}^{K}\left(1-\frac{S \tau_S}{\tau_C}  \right) \log_2(1+{\gamma}_{k}).
\end{equation}

Our goal is to analyze the  ergodic achievable net sum-rate in \eqref{R_sum_MC} above in the large system limit under different CE protocols and channel models, and  study its behaviour versus the network parameters like $M$, $N$ and $K$. We will consider the use of  imperfect I-CSI at the BS to implement precoding, and the use of S-CSI to optimize the RISs phase shifts. The analysis will be done with the objectives of (i) developing optimized RISs designs under different CSI assumptions, (ii) comparing the net sum-rate performance under two different CE protocols for I-CSI acquisition, and (iii) assessing the performance difference between S-CSI and I-CSI based RISs designs.
\vspace{-.05in}

\section{Main Results}
In this section we present our main results starting with preliminaries related to CE, followed by  asymptotic expressions of the net sum-rate under Rician and Rayleigh fading.


\vspace{-.1in}

\subsection{Preliminaries}

\vspace{-.05in}

We first present two CE protocols  used to obtain  I-CSI of $\mathbf{h}_{k}$ to implement precoding.

\subsubsection{MMSE-DFT CE Protocol}  The first CE protocol we consider is the MMSE-DFT protocol from \cite{hiba_CE} which constructs the estimate of the aggregate channel $\mathbf{h}_k$ by estimating $\mathbf{h}_{2lk}$'s and $\mathbf{h}_{dk}$ over $S$  CE sub-phases. In this protocol, RIS $l$ applies the reflect beamforming matrix $\boldsymbol{\Theta}_{ls} = \text{diag}(\phi_{ls1}, \dots, \phi_{lsN}) \in \mathbb{C}^{N \times N}$ in sub-phase $s \in \{1,\dots, S  \}$, resulting in the RISs training matrix  
\begin{align}
\mathbf{V}_{tr}=\begin{bmatrix} 1 & \mathbf{v}_{11}^T & \dots & \mathbf{v}_{L1}^T\\
	\vdots& 	\vdots & &\vdots\\ 
      1 & \mathbf{v}_{1S}^T & \dots & \mathbf{v}_{LS}^T
  \end{bmatrix} \in \mathbb{C}^{S\times (NL+1)} 
  \end{align}
   where $\mathbf{v}_{ls}=\text{diag}(\boldsymbol{\Theta}_{ls}) $.     The optimal $\mathbf{V}_{tr}$ that minimizes the CE error  is derived to be the $NL+1$ leading columns of a DFT matrix as $[\mathbf{V}_{tr}]_{s,n}= e^{-j2\pi(n-1)(s-1)/S }$  in \cite{hiba_CE}.  This protocol  exploits the LoS nature of the BS-RISs channels to reduce the required number of training sub-phases to $S=\frac{NL}{M}+1$  as compared to other CE protocols in \cite{LS, LS1, annie_OJ} which require $S=NL+1$ sub-phases to estimate all channels. This training overhead  is also less than that imposed by  the more recent three-phase and two-phase CE protocols in \cite{CER1}, \cite{CER2}, while achieving a better estimation quality. Moreover this protocol does not  assume any channel sparsity \cite{cas, CE_MU} or consider RIS elements grouping \cite{ofdm_irs, OFDM}.   The  protocol is detailed in \cite{hiba_CE}, and here we extend its results to the channel models in our work, and derive the channel estimates and their statistics as follows, which will be useful to characterize the SINR later on.


\begin{lemma}\label{L11_RIC}
The MMSE estimate of $\mathbf{h}^{\rm ric}_{k}$ in \eqref{eq_ch1_RIC} under the Rician channel models in \eqref{ch111} and the MMSE-DFT CE protocol is given as  \vspace{-.12in}
\begin{align}
\label{est_corr11ric}
\hat{\mathbf{h}}^{\rm ric}_{k}= \hat{\mathbf{h}}^{\rm n}_{dk}+\bar{\mathbf{h}}_{dk}+\sum_{l=1}^L \mathbf{H}_{1l}\boldsymbol{\Theta}_l \hat{\mathbf{h}}^{\rm n}_{2lk}+\sum_{l=1}^L \mathbf{H}_{1l}\boldsymbol{\Theta}_l \bar{\mathbf{h}}_{2lk},
\end{align}
where $\hat{\mathbf{h}}_{dk}^n$ and $\hat{\mathbf{h}}_{2lk}^n$ are the MMSE estimates of $\mathbf{h}^{\rm n}_{dk}$ and $\mathbf{h}^{\rm n}_{2lk}$  given as    \vspace{-.06in}
\begin{align}
\label{h_d_est_RIC}
&\hat{\mathbf{h}}^{\rm n}_{dk}=\frac{\beta^{\rm n}_{dk}}{\beta^{\rm n}_{dk}+\frac{1}{S \rho_{p} \tau_S}}(\tilde{\mathbf{r}}^{tr}_{0k}-\bar{\mathbf{h}}_{dk}), \hspace{.2in} \hat{\mathbf{h}}^{\rm n}_{2lk}=\frac{\beta^{\rm n}_{2lk}}{\beta^{\rm n}_{2lk}+\frac{1}{S \rho_{p} \tau_S M \beta_{1l}} }(\tilde{\mathbf{r}}^{tr}_{lk}-\bar{\mathbf{h}}_{2lk}),
\end{align}
where $\beta^{\rm n}_{dk}=\frac{\beta_{dk}}{\kappa_{dk}+1}$,  $\beta^{\rm n}_{2lk}=\frac{\beta_{2lk}}{\kappa_{2lk}+1}$, and the observation vectors $\tilde{\mathbf{r}}_{0k}^{tr}$ and $\tilde{\mathbf{r}}_{lk}^{tr}$  are given as  \vspace{-.05in}
\begin{align}
\label{MMSE1_direct}
&\tilde{\mathbf{r}}^{tr}_{0k}=\mathbf{h}^{\rm n}_{dk}+\bar{\mathbf{h}}_{dk}+\frac{1}{S} (\mathbf{v}_1^{tr} \otimes \mathbf{I}_{M})^H \mathbf{{n}}^{tr}_{k}, \hspace{.15in} \tilde{\mathbf{r}}^{tr}_{lk}=\mathbf{h}^{\rm n}_{2lk}+\bar{\mathbf{h}}_{2lk}+\frac{1}{SM\beta_{1l}}\bar{\mathbf{H}}_{1l}^H(\mathbf{V}_l^{tr} \otimes \mathbf{I}_{M})^H \mathbf{{n}}^{tr}_{k},
\end{align}
where  $\mathbf{n}_k^{tr} \in \mathbb{C}^{MS \times 1}$ is the received noise across $S$ CE sub-phases, $\rho_p$ is the training SNR, $\tau_S$ is the length of each training sub-phase, $\mathbf{v}_1^{tr}$ is the first $S \times 1$ column of  $\mathbf{V}_{tr}$ and $\mathbf{V}_l^{tr}\in \mathbb{C}^{S \times N}$ comprises of the $N(l-1)+2$ to $Nl+1$ columns of $\mathbf{V}_{tr}$  for $ l=1,\dots, L$. Moreover $\bar{\mathbf{H}}_{1l}= \text{diag}(\bar{\mathbf{h}}_{1l1}, \dots,  \bar{\mathbf{h}}_{1lN}) \hspace{-.04in} \in \hspace{-.04in} \mathbb{C}^{MN \times N}$\hspace{-.04in},  where $\bar{\mathbf{h}}_{1ln}$ is the $n$th column of  $\mathbf{H}_{1l}$, known due to its LoS nature.
\end{lemma}
\begin{IEEEproof}
The proof follows by applying the definition of MMSE estimate on the observation vectors in \eqref{MMSE1_direct} following similar steps as \cite[Sec. III]{hiba_CE}. 
\end{IEEEproof}

Note that under the orthogonality property of MMSE estimates, the CE error $\tilde{\mathbf{h}}^{\rm n}_{dk}=\mathbf{h}^{\rm n}_{dk}-\hat{\mathbf{h}}^{\rm n}_{dk}$, which is also Gaussian, is independent of $\hat{\mathbf{h}}^{\rm n}_{dk}$. A similar discussion applies to  $\tilde{\mathbf{h}}^{\rm n}_{2lk}=\mathbf{h}^{\rm n}_{2lk}-\hat{\mathbf{h}}^{\rm n}_{2lk}$.

Using these results, $\hat{\mathbf{h}}^{\rm ric}_{k}$ is statistically equivalent to a correlated Rician channel as follows.

\begin{lemma} \label{L1_RIC} The channel estimate $\hat{\mathbf{h}}^{\rm ric}_{k}$ in \eqref{est_corr11ric} can be represented as  \vspace{-.12in}
\begin{align}
\label{est_corr_RIC}
& \hat{\mathbf{h}}^{\rm ric}_{k}= \bar{\mathbf{h}}_{dk}+\sum_{l=1}^L \mathbf{H}_{1l}\boldsymbol{\Theta}_l \bar{\mathbf{h}}_{2lk}+\mathbf{C}^{\rm ric^{1/2}}_{k}\mathbf{q}_{k},
\end{align}
where $\mathbf{q}_{k}\sim \mathcal{CN}(0,\mathbf{I}_{M})$ and $\mathbf{C}^{\rm ric}_{k}=\frac{\beta_{dk}^{{n}^{2}}}{\beta^{\rm n}_{dk}+\frac{1}{S\rho_{p} \tau_S}} \mathbf{I}_M+\sum_{l=1}^L \frac{\beta_{2lk}^{{n}^{2}}}{\beta^{\rm n}_{2lk}+\frac{1}{S\rho_{p} \tau_S M \beta_{1l}}} \mathbf{H}_{1l}  \mathbf{H}_{1l}^H$.
\end{lemma}
\begin{IEEEproof}
The proof is provided in Appendix \ref{appena4}.
\end{IEEEproof}

For Rayleigh fading, the following corollaries can be obtained from Lemma \ref{L11_RIC} and Lemma \ref{L1_RIC}.
\begin{corollary} \label{L00} The MMSE estimate of $\mathbf{h}_k^{\rm ray}=\mathbf{h}_{dk}^{\rm ray}+\sum_{l=1}^L\mathbf{H}_{1l}\boldsymbol{\Theta}_l\mathbf{h}_{2lk}^{\rm ray}$ under Rayleigh fading $\mathbf{h}_{dk}$'s and $\mathbf{h}_{2lk}$'s and the MMSE-DFT CE protocol is given as 
\begin{align}
\label{est_corr11}
\hat{\mathbf{h}}^{\rm ray}_{k}= \hat{\mathbf{h}}^{\rm ray}_{dk}+\sum_{l=1}^L \mathbf{H}_{1l}\boldsymbol{\Theta}_l \hat{\mathbf{h}}^{\rm ray}_{2lk},
\end{align}
where the MMSE estimates of $\mathbf{h}^{\rm ray}_{dk}$ and ${\mathbf{h}}^{\rm ray}_{2lk}$ are $\hat{\mathbf{h}}^{\rm ray}_{dk}=\frac{\beta_{dk}}{\beta_{dk}+\frac{1}{S\rho_{p} \tau_S}}\tilde{\mathbf{r}}^{tr}_{0k}$, $\hat{\mathbf{h}}^{\rm ray}_{2lk}= \frac{\beta_{2lk}}{\beta_{2lk}+\frac{1}{S\rho_{p} \tau_S M \beta_{1l}}}\tilde{\mathbf{r}}^{tr}_{lk}$, 
where $\tilde{\mathbf{r}}^{tr}_{0k}=\mathbf{h}^{\rm ray}_{dk}+\frac{1}{S} (\mathbf{v}_1^{tr} \otimes \mathbf{I}_{M})^H \mathbf{{n}}^{tr}_{k}$, and  $\tilde{\mathbf{r}}^{tr}_{lk}= \mathbf{h}^{\rm ray}_{2lk}+\frac{1}{SM\beta_{1l}}\bar{\mathbf{H}}_{1l}^H(\mathbf{V}_l^{tr} \otimes \mathbf{I}_{M})^H \mathbf{{n}}^{tr}_{k}$.
\end{corollary}

\begin{corollary} \label{L1}
The channel estimate $\hat{\mathbf{h}}^{\rm ray}_{k}$ in \eqref{est_corr11}  is statistically equivalent to
\begin{align}
\label{est_corr}
& \hat{\mathbf{h}}^{\rm ray}_{k}= \mathbf{C}_{k}^{{\rm ray}^{1/2}}\mathbf{q}_{k},
\end{align}
where $\mathbf{q}_{k}\sim \mathcal{CN}(0,\mathbf{I}_{M})$ and $\mathbf{C}^{\rm ray}_{k}=\frac{\beta_{dk}^{2}}{\beta_{dk}+\frac{1}{S\rho_{p} \tau_S}} \mathbf{I}_M+\sum_{l=1}^L \frac{\beta_{2lk}^{2}}{\beta_{2lk}+\frac{1}{S\rho_{p} \tau_S M \beta_{1l}}}  \mathbf{H}_{1l} \mathbf{H}_{1l}^H$.
\end{corollary}
\begin{IEEEproof}
These results follow by setting $\kappa_{2lk}$ and $\kappa_{dk}$ to $0$ in Lemma \ref{L11_RIC} and \ref{L1_RIC}.
\end{IEEEproof}

The MMSE-DFT  protocol estimates all channels accurately at the expense of a large overhead which may compromise the  net sum-rate. Next, we present a lower overhead CE protocol.

\subsubsection{DE Protocol}

In the DE scheme, instead of estimating the individual channels $\mathbf{h}_{dk}$'s and $\mathbf{h}_{2lk}$'s, the BS directly estimates   the aggregate channel $\mathbf{h}_{k}=\mathbf{h}_{dk}+\sum_{l=1}^L \mathbf{H}_{1l}\boldsymbol{\Theta}_l\mathbf{h}_{2lk}$ for given RISs  beamforming matrices $\boldsymbol{\Theta}_l$'s in a single sub-phase. The received training signal is correlated with each user's pilot sequence to obtain the received  observation vectors $\mathbf{y}^{tr}_{k}= \mathbf{h}_{k} +\mathbf{n}_{k}^{tr}, \hspace{.02in} k=1,\dots, K$, where $\mathbf{n}_{k}\sim \mathcal{CN}(\mathbf{0}, \frac{1}{\rho_p \tau_S}\mathbf{I}_M)$, that are used to estimate $\mathbf{h}_k$. Since  DE is done for given $\boldsymbol{\Theta}_l$'s, the channel estimate of $\mathbf{h}_k$ will depend on the choice of RIS phase shifts. We consider RISs to implement the same $\boldsymbol{\Theta}_l$'s over multiple CE and downlink data transmission phases, that are found by optimizing the ergodic net sum-rate performance based on S-CSI. In terms of implementation, we  define a time-frame consisting of several coherence periods over which the channel statistics stay constant, where each coherence period is further divided into a CE phase and a downlink transmission phase. When the time-frame starts the BS uses its knowledge of channel statistics to find   $\boldsymbol{\Theta}_l$'s that maximize the net sum-rate (as discussed in the next section), which are then used during the CE phase (in which $\mathbf{h}_{k}$'s are estimated) as well as during the downlink transmission phase of each coherence interval in that time-frame. Once the time-frame ends, the channel statistics are re-acquired and $\boldsymbol{\Theta}_l$'s are re-computed for the next time frame.

The MMSE estimate of $\mathbf{h}_{k}$ under DE, for  given $\boldsymbol{\Theta}_l$'s, is  stated next for both fading models.
\begin{lemma} \label{L4ric}  The MMSE estimate of $\mathbf{h}^{\rm ric}_{k}$  \eqref{eq_ch1_RIC} under  Rician fading and  DE protocol is given as 
\begin{align}
\label{est_deric}
& \hat{\mathbf{h}}^{\rm ric}_{k}= \bar{\mathbf{h}}_{dk}+\sum_{l=1}^L\mathbf{H}_{1l}\boldsymbol{\Theta}_l\bar{\mathbf{h}}_{2lk}+\mathbf{R}^{\rm ric}_{k} \mathbf{Q}^{\rm ric}_{k} (\mathbf{y}_{k}^{tr}-\bar{\mathbf{h}}_{dk}-\sum_{l=1}^L\mathbf{H}_{1l}\boldsymbol{\Theta}_l\bar{\mathbf{h}}_{2lk}),
\end{align}
where $\mathbf{y}_{k}^{tr}=\mathbf{h}_{dk}^{\rm ric}+\sum_{l=1}^L\mathbf{H}_{1l}\boldsymbol{\Theta}_l{\mathbf{h}}^{\rm ric}_{2lk}+\mathbf{n}_k^{tr}$, $\mathbf{Q}^{\rm ric}_{k}=\left(\mathbf{R}^{\rm ric}_{k}+\frac{ \mathbf{I}_M}{\rho_{p} \tau_S} \right)^{-1}$, and $\mathbf{R}^{\rm ric}_{k}=\beta^{\rm n}_{dk}\mathbf{I}_M +\sum_{l=1}^L\beta^{\rm n}_{2lk} \mathbf{H}_{1l} \mathbf{H}_{1l}^H$. The channel estimate $\hat{\mathbf{h}}^{\rm ric}_{k}$  is statistically equivalent to $ \hat{\mathbf{h}}^{\rm ric}_{k}= \mathbf{C}_{k}^{{\rm DE, ric}^{1/2}}\mathbf{q}_{k}$, where $\mathbf{q}_{k}\sim \mathcal{CN}(0,\mathbf{I}_{M})$ and $\mathbf{C}^{\rm DE, ric}_{k}=\mathbf{R}^{\rm ric}_{k} \mathbf{Q}^{\rm ric}_{k}\mathbf{R}^{\rm ric}_{k} $.
\end{lemma}
 \begin{IEEEproof}
The proof follows from writing $\mathbf{h}_k^{\rm ric}$ as a sum of LoS and NLoS channels and estimating the NLoS part $\mathbf{h}_{k}^{\rm n}=\mathbf{h}_{dk}^{\rm n}+\sum_{l=1}^L\mathbf{H}_{1l}\boldsymbol{\Theta}_l{\mathbf{h}}^{\rm n}_{2lk}$  using $\mathbf{y}_{k}^{tr}-\bar{\mathbf{h}}_{dk}-\sum_{l=1}^L\mathbf{H}_{1l}\boldsymbol{\Theta}_l\bar{\mathbf{h}}_{2lk}$.
\end{IEEEproof}

\begin{corollary} \label{L4} The MMSE estimate of  $\mathbf{h}_k^{\rm ray}$ under  Rayleigh fading and DE protocol is given as
\begin{align}
\label{est_de}
& \hat{\mathbf{h}}^{\rm ray}_{k}= \mathbf{R}^{\rm ray}_{k} \mathbf{Q}^{\rm ray}_{k} \mathbf{y}_{k}^{tr},
\end{align}
where $\mathbf{R}^{\rm ray}_{k}=\beta_{dk}\mathbf{I}_M +\sum_{l=1}^L\beta_{2lk} \mathbf{H}_{1l} \mathbf{H}_{1l}^H$, $\mathbf{Q}^{\rm ray}_{k}=\left(\mathbf{R}^{\rm ray}_{k}+\frac{ \mathbf{I}_M}{\rho_{p}\tau_S} \right)^{-1}$, and $\mathbf{y}_{k}^{tr}=\mathbf{h}_{dk}^{\rm ray} \\ +\sum_{l=1}^L\mathbf{H}_{1l} \boldsymbol{\Theta}_l \mathbf{h}_{2lk}^{\rm ray}+\mathbf{n}_k^{tr}$. The channel estimate $\hat{\mathbf{h}}^{\rm ray}_{k}$  is statistically equivalent to $ \hat{\mathbf{h}}^{\rm ray}_{k}= \mathbf{C}_{k}^{{\rm DE, ray}^{1/2}}\mathbf{q}_{k}$, where $\mathbf{q}_{k}\sim \mathcal{CN}(0,\mathbf{I}_{M})$ and $\mathbf{C}^{\rm DE, ray}_{k}=\mathbf{R}^{\rm ray}_{k} \mathbf{Q}^{\rm ray}_{k}\mathbf{R}^{\rm ray}_{k} $.
 \end{corollary}
 \begin{IEEEproof}
The proof follows from setting $\kappa_{dk}$ and $\kappa_{2lk}$ as $0$ in Lemma \ref{L4ric}.
\end{IEEEproof}

While this protocol does not provide full I-CSI of the individual RIS-assisted links, it provides enough information to implement precoding at the BS, which only requires the estimate of the aggregate channel $\hat{\mathbf{h}}_k$. It also saves the large training overhead associated with $S\geq \frac{NL}{M}+1$ sub-phases  to obtain the full I-CSI under MMSE-DFT protocol, and requires only a single training sub-phase.  The downside is that  the estimates in \eqref{est_deric} and \eqref{est_de}  can not be used to design the RIS phases instantaneously as that will require the estimates $\hat{\mathbf{h}}_{dk}$'s and $\hat{\mathbf{h}}_{2lk}$'s. However, if the RISs phases are designed using  S-CSI which we  will investigate next, then DE is a  desirable scheme because the BS can use \eqref{est_deric} and \eqref{est_de} instead of \eqref{est_corr11ric} and \eqref{est_corr11} to implement precoding.


\subsection{Asymptotic Analysis under Rician Fading}

 While the users' ergodic rates in \eqref{rate_11} are generally difficult to study for finite system dimensions, they  tend to approach deterministic quantities as the system dimensions grow large, which are referred to as deterministic equivalents. While these deterministic equivalents are almost surely (a.s.) tight in the asymptotic limit, they are very accurate for moderate system dimensions as well as observed in \cite{HJY, ML2, annie, jia, wag33}. Consequently, they can be used to gain insights into the behaviour of the system when  $M$ and $K$ are finite. Moreover, they depend only on the channel statistics and are very useful to solve important optimization problems in  massive MIMO literature based on S-CSI.  Under this motivation, we exploit the statistical distribution of $\mathbf{h}_{k}$ and large values of $M,N,K$ envisioned for beyond 5G  networks to compute the deterministic approximations of the users' ergodic rates under the following required assumptions  \cite{HJY, annie}.

\textit{ Assumption 1.}
$M$, $N$ and $K$ grow large with a bounded ratio as $0< \liminf_{M,K \rightarrow \infty} \frac{K}{M}\leq \limsup_{M,K \rightarrow \infty}\frac{K}{M}<\infty$ and $0< \liminf_{M,N \rightarrow \infty} \frac{M}{N}\leq \limsup_{M,N \rightarrow \infty}\frac{M}{N}<\infty$. 

\textit{ Assumption 2.}
The LoS channel matrix $\mathbf{H}_1$ satisfies $\limsup\limits_{M,N\rightarrow \infty} ||\mathbf{H}_1\mathbf{H}_1^{H}||<\infty$.



Now we are ready to present the asymptotic analysis  under the two considered channel models and CE protocols.  We first present the deterministic equivalents of the SINR under Rician fading.

\begin{theorem} \label{thm1_ric} Under Assumptions 1 and 2, the SINR of user $k$  in (\ref{SINR_MRT}), for the channel in (\ref{eq_ch1_RIC}) and its estimate in (\ref{est_corr_RIC}) under the MMSE-DFT  protocol satisfies $\gamma_{k}^{\rm ric}-{\gamma}_{k}^{\rm ric^\circ }\xrightarrow[M,N,K\rightarrow \infty]{a.s} 0$,  where \vspace{-.1in}
\begin{align}
\label{det_SINR_RIC}
&{\gamma}_{k}^{\rm ric^\circ}=\frac{p_k\left|\frac{1}{M}\text{tr}\left(\mathbf{D}_{k}+\sum_{l=1}^{L}\frac{\beta_{2lk}^{{n}^2}}{\beta_{2lk}^{\rm n}+\frac{1}{S\rho_{p}\tau_S M \beta_{1l}}} \mathbf{H}_{1l}\mathbf{H}_{1l}^H+ \frac{\beta_{dk}^{{n}^2}}{\beta_{dk}^{\rm n}+\frac{1}{S\rho_{p}\tau_S}} \boldsymbol{I}_{M}\right)\right|^{2}}{\frac{1}{M}\sum_{f\neq k}{\frac{p_{f}}{M}\text{tr}((\mathbf{D}_{f}+\textbf{C}_{f}^{\rm ric})(\mathbf{D}_{k}+\mathbf{A}_{k}^{\rm ric})})+\frac{\frac{p_{k}}{M}\sum_{k=1}^{K}{\frac{1}{M}\text{tr}(\mathbf{D}_{k}+\mathbf{C}_{k}^{\rm ric})}}{\rho}},
\end{align}
$\mathbf{D}_{k}=\bar{\mathbf{h}}_{dk}\bar{\mathbf{h}}^{H}_{dk}+\bar{\mathbf{h}}_{dk}\sum_{l=1}^{L}\bar{\mathbf{h}}_{2lk}^{{H}} \boldsymbol{\Theta}_l^H \mathbf{H}_{1l}^H+\sum_{l=1}^{L}\mathbf{H}_{1l} \boldsymbol{\Theta}_l \bar{\mathbf{h}}_{2lk} \bar{\mathbf{h}}_{dk}^{H}+\sum_{l=1}^{L}\sum_{l'=1}^{L}\mathbf{H}_{1l} \boldsymbol{\Theta}_l \bar{\mathbf{h}}_{2lk}\bar{\mathbf{h}}_{2l'k}^{{H}} \boldsymbol{\Theta}_{l'}^H \mathbf{H}_{1l'}^H$, $\mathbf{A}_{k}^{\rm ric}$ is defined in \eqref{eq_ch1_RIC22} and $\textbf{C}_{k}^{\rm ric}$ is defined in Lemma \ref{L1_RIC}.

\end{theorem}
\begin{IEEEproof}
The proof of Theorem \ref{thm1_ric} is provided in Appendix \ref{appenb_ric}.
\end{IEEEproof}


\begin{theorem}\label{thm2_ric} 
  Under Assumptions 1 and 2, the SINR of user $k$  in (\ref{SINR_MRT}), for the channel in (\ref{eq_ch1_RIC}) and its estimate in \eqref{est_deric} under the MMSE-DE protocol, satisfies  $\gamma_{k}^{\rm ric}-{\gamma}_{k}^{\rm ric^\circ}\xrightarrow[M,N,K\rightarrow \infty]{a.s} 0$, 	where  \vspace{-.1in}
\begin{align}
\label{det_SINR_DE_ric}
&\gamma_{k}^{\rm ric^\circ}=\frac{p_{k}|\frac{1}{M}\text{tr}(\mathbf{D}_{k}+\mathbf{R}^{\rm ric}_{k} \mathbf{Q}^{\rm ric}_{k}\mathbf{R}_{k}^{\rm ric})|^2}{\frac{1}{M}\sum_{f\neq k} \frac{p_{f}}{M}\text{tr}((\mathbf{D}_{f}+\mathbf{R}^{\rm ric}_{f} \mathbf{Q}^{\rm ric}_{f}\mathbf{R}^{\rm ric}_{f})( \mathbf{D}_{k}+\mathbf{R}^{\rm ric}_{k}))+\frac{1}{M} \sum_{k=1}^{K}\frac{\frac{p_{k}}{M} \text{tr}(\mathbf{D}_{k}+\mathbf{R}^{\rm ric}_{k} \mathbf{Q}^{\rm ric}_{k} \mathbf{R}^{\rm ric}_{k})}{\rho}},
\end{align} 
where $\mathbf{R}^{\rm ric}_{k}$ and $\mathbf{Q}^{\rm ric}_{k}$ are defined in Lemma \ref{L4ric}, and $\mathbf{D}_k$ is defined in Theorem \ref{thm1_ric}.
\end{theorem}
\begin{IEEEproof}
The proof of Theorem \ref{thm2_ric} is similar to the proof of Theorem \ref{thm1_ric} while considering the MMSE channel estimates  defined in Lemma \ref{L4ric}.
\end{IEEEproof}
Next, we simplify the result under perfect CSI, which will be used in the sequel for comparison.

\begin{corollary}\label{cor_perric} Under the setting of Theorem \ref{thm1_ric}, and assuming perfect CSI , $\gamma_k^{\rm ric^{\circ}}$ is given as 
\begin{equation}
\label{det_Ak_RIC}
{\gamma}_{k}^{\rm ric^\circ}=\frac{p_{k}|\frac{1}{M}\text{tr}(\mathbf{D}_{k}+\mathbf{A}_{k}^{\rm ric})|^{2}}{\frac{1}{M}\sum_{f\neq k}\frac{p_{f}}{M}{\text{tr}((\mathbf{D}_{f}+\mathbf{A}_{f}^{\rm ric})(\mathbf{D}_{k}+\mathbf{A}_{k}^{\rm ric}}))+\frac{p_{k}}{M\rho}\sum_{k=1}^{K}{\frac{1}{M}\text{tr}(\mathbf{D}_{k}+\mathbf{A}_{k}^{\rm ric})}}.
\end{equation}
\end{corollary}
\begin{IEEEproof}
The proof follows by letting $\rho_{p}\rightarrow \infty$ in Theorem \ref {thm1_ric}.
\end{IEEEproof}

We also simplify the result in Theorem \ref{thm2_ric} for the scenario without RISs for comparison.
\begin{theorem}\label{thm2_ricwo} 
  Consider the setting of Theorem \ref{thm2_ric} without RISs, then ${\gamma}_{k}^{\rm ric^\circ}$ is given as \small
\begin{align}
\label{det_SINR_DE_ricwo}
&\gamma_{k}^{\rm ric^\circ}\hspace{-.08in}=\hspace{-.05in}\frac{p_{k}\left|\frac{1}{M}\left(\bar{\mathbf{h}}_{dk}^H \bar{\mathbf{h}}_{dk} +\frac{\beta_{dk}^{{n}^2}}{\beta_{dk}^{\rm n}+\frac{1}{\rho_{p}\tau_S}}M\right)\right|^2}{\frac{1}{M}\sum_{f\neq k} \frac{p_{f}}{M}\text{tr}\left(\left(\bar{\mathbf{h}}_{df} \bar{\mathbf{h}}^H_{df} +\frac{\beta_{df}^{{n}^2}\mathbf{I}_M}{\beta_{df}^{\rm n}+\frac{1}{\rho_{p}\tau_S}}\right)\left(\bar{\mathbf{h}}_{dk} \bar{\mathbf{h}}^H_{dk} +\frac{\beta_{dk}^{{n}^2}\mathbf{I}_M}{\beta_{dk}^{\rm n}+\frac{1}{\rho_{p}\tau_S}}\right)\right)+\frac{1}{M} \sum_{k=1}^{K}\frac{p_{k}}{M\rho}\left(\bar{\mathbf{h}}_{dk}^H \bar{\mathbf{h}}_{dk} +\frac{\beta_{dk}^{{n}^2}M}{\beta_{dk}^{\rm n}+\frac{1}{\rho_{p}\tau_S}}\right)}.
\end{align} \normalsize
\end{theorem}
\begin{IEEEproof}
The proof of Theorem \ref{thm2_ricwo} follows by setting $N$ and $L$ as zero in Theorem \ref{thm2_ric}.
\end{IEEEproof}

Using these results, we can obtain the  deterministic equivalents of ergodic net rates as follows.
\begin{corollary} \label{Cor_rate_RIC}
Under Assumptions 1 and 2,  the users' ergodic achievable net rates in \eqref{rate_11} under Rician fading, denoted by $R^{\rm ric}_{k}$, converge as $R^{\rm ric}_{k}-R_{k}^{\rm ric^\circ}\xrightarrow[M,N,K\rightarrow \infty]{a.s} 0$,  where $R_{k}^{\rm ric^\circ}= \left(1-\frac{S \tau_S}{\tau_C}  \right)\log(1+{\gamma}_{k}^{\rm ric^\circ})$ and ${\gamma}_{k}^{\rm ric^\circ}$ is given by (\ref{det_SINR_RIC}) with $S=\frac{NL}{M}+1$ for the MMSE-DFT CE protocol, or given by \eqref{det_SINR_DE_ric} with $S=1$ for the MMSE-DE CE protocol.
\end{corollary}
\begin{IEEEproof}
The proof follows by applying the continuous mapping theorem  on $R_k^{\rm ric}$.
\end{IEEEproof}

 An asymptotic approximation for the ergodic achievable  net sum-rate can be obtained as \vspace{-.1in}
\begin{equation}
\label{R_sum_ric}
R_{sum}^{\rm ric^\circ}= \sum_{k=1}^{K}\left(1-\frac{S \tau_S}{\tau_C}  \right) \log(1+{\gamma}_{k}^{\rm ric^\circ}).
\end{equation}

 
The deterministic equivalents in \eqref{det_SINR_RIC}, \eqref{det_SINR_DE_ric} and \eqref{det_SINR_DE_ricwo} provide some insights into the  behaviour of massive MIMO  under Rician fading. The desired signal energy in the numerator of  these expressions  stays constant with respect to $M$ since each term is a ratio  of the trace of an $M\times M$ matrix to  $M$. On the other hand, following \cite{ablaric}, the interference term and noise term in the denominators of all three expressions vanish  as $M\rightarrow \infty$ while $K, N$ and $L$ are kept fixed. Therefore, the SINR grows with $M$ for fixed $K$, which is referred to as the ``massive MIMO" effect in \cite{HJY}. We also note from  \eqref{det_SINR_RIC} and \eqref{det_SINR_DE_ric} that asymptotically, the RIS reflect beamforming matrices $\boldsymbol{\Theta}_l$'s appear in  all terms involving the LoS channel components and do not appear  in the NLoS terms. Therefore RISs will yield higher performance gains for large  Rician factors $\kappa$. Moreover, the desired signal  and interference  (first term in denominator) terms scale quadratically with $N$ and $L$, while the noise term scales linearly with $N$ and $L$, when considering fixed RISs phases indicating more RIS gains in noise limited scenarios. However, we can optimize the  phase shifts to improve  SINR  significantly in interference limited scenarios. 

\subsection{Analysis under Rayleigh Fading}
While  RISs provide beamforming gains under Rician fading, we will see from the deterministic equivalents presented next that they  only provide an array gain under Rayleigh fading. 
\begin{corollary} \label{th2} Under Assumptions 1 and 2, the SINR of user $k$ defined in (\ref{SINR_MRT}), for the channel  estimate in (\ref{est_corr}) under the MMSE-DFT CE protocol satisfies $\gamma_{k}^{\rm ray}-{\gamma}_{k}^{\rm ray^\circ }\xrightarrow[M,N,K\rightarrow \infty]{a.s} 0$,  where
\begin{align}
\label{det_SINR}
{\gamma}_{k}^{\rm ray ^\circ } = \frac{p_{k}\left|\frac{1}{M}\text{tr}\left(\sum_{l=1}^{L}\frac{\beta_{2lk}^{2}}{\beta_{2lk}+\frac{1}{S\rho_{p}\tau_S M \beta_{1l}}} \mathbf{H}_{1l}\mathbf{H}_{1l}^H+ \frac{\beta_{dk}^{2}}{\beta_{dk}+\frac{1}{S\rho_{p}\tau_S}} \boldsymbol{I}_{M}\right)\right|^{2}}{\frac{1}{M}\sum_{f\neq k}{\frac{p_{f}}{M}\text{tr}(\textbf{C}_{f}^{\rm ray}\mathbf{A}_{k}^{\rm ray})}+\frac{\frac{p_{k}}{M}\sum_{k=1}^{K}{\frac{1}{M} \text{tr}(\mathbf{C}_{k}^{\rm ray})}}{\rho}}
\end{align}
where $\mathbf{A}_k^{\rm ray}=\sum_{l=1}^L \beta_{2lk}\mathbf{H}_{1l}\mathbf{H}_{1l}^H+\beta_{dk}\mathbf{I}_M$ and $\textbf{C}^{\rm ray}_{k}$ is defined in Corollary 2.
\end{corollary}
\begin{IEEEproof}
The proof follows by setting $\kappa_{dk}$ and $\kappa_{2lk}$ as $0$ in Theorem \ref{thm1_ric}.
\end{IEEEproof}

\begin{corollary}\label{Thm2} 
  Under Assumptions 1 and 2, the SINR of user $k$ defined in (\ref{SINR_MRT}), for the channel estimate  in \eqref{est_de} under the MMSE-DE protocol, satisfies $	\gamma^{\rm ray}_{k}-{\gamma}_{k}^{\rm ray^\circ }\xrightarrow[M,N,K\rightarrow \infty]{a.s} 0$, 	where \vspace{-.07in}
\begin{align}
\label{det_SINR_DE}
&\gamma_{k}^{\rm ray^\circ }=\frac{p_{k}\left|\frac{1}{M}\text{tr}(\mathbf{R}^{\rm ray}_{k} \mathbf{R}^{\rm ray}_{k}\mathbf{Q}^{\rm ray}_{k})\right|^2}{\frac{1}{M}\sum_{f\neq k} \frac{p_{f}}{M}\text{tr}(\mathbf{R}^{\rm ray}_{k} \mathbf{R}^{\rm ray}_{f} \mathbf{Q}^{\rm ray}_{f} \mathbf{R}^{\rm ray}_{f})+\frac{1}{M}\sum_{k=1}^{K}\frac{\frac{p_{k}}{M} \text{tr}(\mathbf{R}^{\rm ray}_{k} \mathbf{R}^{\rm ray}_{k} \mathbf{Q}^{\rm ray}_{k})}{\rho}},
\end{align} 
and $\mathbf{R}^{\rm ray}_{k}$ and $\mathbf{Q}^{\rm ray}_{k}$ are defined in Corollary 3.
\end{corollary}
\begin{IEEEproof}
The proof is obtained by setting $\kappa_{dk}$ and $\kappa_{2lk}$ as $0$ in Theorem 2.
\end{IEEEproof}

\begin{corollary} \label{Cor_per} Under the setting of Corollary \ref{th2} and \ref{Thm2} and assuming perfect CSI, we have
\begin{equation}
\label{ray_per}
{\gamma}_{k}^{\rm ray^{\circ} }=\frac{p_{k}|\frac{1}{M}\text{tr}(\mathbf{A}_{k}^{\rm ray})|^{2}}{\frac{1}{M}\sum_{f\neq k}{\frac{p_{f}}{M}\text{tr}(\mathbf{A}_{f}^{\rm ray}\mathbf{A}_{k}^{\rm ray}})+\frac{\frac{p_{k}}{M}\sum_{k=1}^{K}{\frac{1}{M}\text{tr}(\mathbf{A}_{k}^{\rm ray})}}{\rho}}.
\end{equation}
\end{corollary}
\begin{IEEEproof}
The proof follows by letting $\rho_{tr}\rightarrow \infty$ in Corollary \ref{th2} and \ref{Thm2}.
\end{IEEEproof}

Next we express the deterministic equivalents of users' ergodic rates in the following corollary.
\begin{corollary} \label{Cor_rate}
Under Assumptions 1 and 2, the users' ergodic  net rates under Rayleigh fading, denoted by  $R^{\rm ray}_{k}$, converge as $R^{\rm ray}_{k}-R_{k}^{\rm ray^\circ }\xrightarrow[M,N,K\rightarrow \infty]{a.s} 0$,  where $R_{k}^{\rm ray^\circ }= \left(1-\frac{S\tau_S}{\tau_C}\right)\log(1+{\gamma}_{k}^{\rm ray^\circ })$ and ${\gamma}_{k}^{\rm ray^\circ }$ is given by (\ref{det_SINR}) with $S=\frac{NL}{M}+1$ for the MMSE-DFT protocol, and given by \eqref{det_SINR_DE} with $S=1$ for the MMSE-DE protocol. The ergodic achievable net sum-rate is given as \vspace{-.06in}
\begin{equation}
\label{R_sum_ray}
R_{sum}^{\rm ray^\circ }= \sum_{k=1}^{K}\left(1-\frac{S\tau_S}{\tau_C}\right) \log(1+{\gamma}_{k}^{\rm ray^\circ }).
\end{equation}
\end{corollary}
\begin{IEEEproof}
The proof is similar to the one for Corollary 5.
\end{IEEEproof}


Two important insights can be drawn from these expressions. First, the phase-shifts applied by the RISs, i.e. $\phi_{ln}=\exp(j\theta_{ln})$, do not appear in the deterministic equivalents of the SINR under Rayleigh fading, implying that the RISs will not yield any reflect beamforming gain. This phenomenon also observed in \cite{IRSstat1} is caused by the spatial isotropy that holds upon the RIS-assisted channel, which is insensitive to the beamforming between $\mathbf{H}_1$ and $\mathbf{h}_{2lk}^{\rm ray}$ under Rayleigh fading $\mathbf{h}_{2lk}^{\rm ray}$. Second, the RISs can still yield an array gain due to the sum over $N$ and $L$ terms in the numerator. However, there are also terms with sum over $N$ and $L$ in the denominator. We will draw insights next as to when is deploying  RISs useful under independent Rayleigh fading.

%

\textit{How Useful is the RIS under Rayleigh Fading?:} To gain explicit insights into the impact of RISs on the SINR under Rayleigh fading, we consider a special case which assumes  \vspace{-.08in}
\begin{align}
\label{spec}
&\mathbf{H}_{1l}=\sqrt{\beta_{1l} N}\textbf{U}_l,
\end{align}
 where $\textbf{U}_l\in \mathbb{C}^{M\times N}$ is composed of $M\leq N$ leading rows of an arbitrary $N\times N$ unitary matrix \cite{HJY}. Since in practice each diagonal entry of $\mathbf{H}_{1l}\mathbf{H}_{1l}^H$ is the sum of $N$ exponential terms of unit norm as can be seen from \eqref{H_1}, so we have normalized $\mathbf{U}_l$ by  $\sqrt{N}$. Moreover $N$ is assumed to be large but fixed to ensure a bounded spectral norm. The model implies that $\mathbf{H}_{1l}$ has orthogonal rows. Such a LoS scenario can be realized in practice through specific placement of the RISs with respect to the BS array. This special case will act as an upper-bound on the RIS performance under arbitrary $\mathbf{H}_{1l}$'s. Moreover, we let $\beta_{1l} \beta_{2lk}=c_l\beta_{dk}$, $\forall l,k$, that is justified in scenarios where each RIS is located  close to the BS. For this special case, the performance of the RISs-assisted system under perfect CSI is given in a compact closed-form as follows.

\begin{corollary}\label{Cor:spec} For the special case in \eqref{spec}, ${\gamma}_{k}^{\rm ray^\circ }$ in Corollary \ref{Cor_per} under  $p_k=1$, $\forall k$ is given as \vspace{-.05in}
\begin{align}
\label{special1}
&{\gamma}_{k}^{\rm ray^\circ }=\frac{1}{\underbrace{\frac{1}{M}\sum_{f\neq k} \frac{\beta_{df}}{\beta_{dk}}}_{\text{Interference}}+\underbrace{\frac{\sum_{k'
=1}^K \beta_{dk'}}{M \beta_{dk}^2\rho (\bar{c}N+1)}}_{\text{Noise}}},
\end{align}
\end{corollary}\vspace{-.35in}
where $\bar{c}=\sum_{l=1}^L c_l$. 
\begin{IEEEproof}
The proof follows by  substituting \eqref{spec} for $\mathbf{H}_{1l}$'s and $\beta_{1l} \beta_{2lk}=c_l\beta_{dk}$ in \eqref{ray_per}.
\end{IEEEproof}

This corollary yields two important insights. First it verifies the ``massive MIMO effect" observed in \cite{HJY}, that the SINR increases with $M$  for fixed $N$, $L$, and $K$. Second, the use of  RISs under Rayleigh fading $\mathbf{h}_{2lk}$'s is only useful in large systems when the average received SNR is low, i.e. either $\rho$ is low or the path loss is high. This is often the case for cell-edge users. In such a noise-limited scenario, the second term in the denominator of \eqref{special1} will dominate the first and increasing $N$ and $L$ will produce a noticeable increase in the SINR. In an interference-limited scenario, the use of  RISs yields no substantial benefit. This is because  under Rayleigh fading, each RIS  yields an array gain of $N$ asymptotically, which appears in both the energy of desired and interfering signals and the net effect becomes negligible if   interference is dominant. 

\section{Performance Optimization} \vspace{-.05in}

Since in the large system limit, the RISs yield reflect beamforming gains under Rician fading  only, therefore we focus on the design of RISs phase shifts for Rician fading channels.
\subsection{Ergodic Net  Sum-Rate Maximization using S-CSI}

In this section, we optimize the RIS phase-shifts by utilizing the deterministic equivalents of the  ergodic net sum-rate  derived in Sec. III-B in \eqref{R_sum_ric}. This approach allows us to optimize the RISs phase shifts using knowledge of only the large-scale channel statistics that characterize these deterministic equivalents. This S-CSI changes much slower than the actual fast fading channel itself, and therefore the RISs phase shifts need to be optimized only once after several coherence periods which reduces complexity. The net sum-rate maximization problem is formulated below. \vspace{-.25in}
 \begin{subequations}
 \begin{alignat}{2} \textit{(P1)} \hspace{.15in}
&\!\max_{\boldsymbol{\phi}} \hspace{.25in} &  \sum_{k=1}^{K} \left(1-\frac{S \tau_S}{\tau_C}  \right)  \log(1+\gamma_{k}^{\rm ric^\circ}) \label{P1}\\
&\text{s.t.} & \hspace{.25in} |\phi_{ln}|=1, \hspace{.08in} \forall l,n.\label{constraint7}
\end{alignat}
\end{subequations} 
where $\phi_{ln}$ is the $n^{th}$ diagonal element of $\boldsymbol{\Theta}_l$, and  $\boldsymbol{\phi}=[\phi_{11}, \dots, \phi_{1N}, \phi_{21}, \dots, \phi_{LN}]^T\in \mathbb{C}^{NL\times 1}$.


\begin{algorithm}[!t]
\caption{Projected Gradient Ascent Algorithm for the RISs Design}\label{alg:euclid}
\begin{algorithmic}[1]
\State \textbf{Initialize:} $\boldsymbol{\phi}^1$, $R_{sum}^{\rm ric^{\circ^1}}=f(\boldsymbol{\phi}^1)$ where $f(.)$ is given by \eqref{R_sum_ric}, $\epsilon>0$, $s=1$.
\State  \textbf{Repeat}
\State $\bar{R}_{sum}^{\rm ric^{\circ}}=R_{sum}^{\rm ric^{\circ^s}}$;
\State $[\mathbf{p}^{s}]_{N(l-1)+n}=\frac{\partial R_{sum}^{\rm ric^\circ}}{\partial \phi_{ln}}|_{\boldsymbol{\phi}^s}$, $n=1,\dots, N$, $l=1,\dots, L$;
\State $\tilde{\boldsymbol{\phi}}^{s+1}=\boldsymbol{\phi}^{s}+\mu \mathbf{p}^{s}$; 
\State  $\boldsymbol{\phi}^{s+1}=\exp(j \text{arg } (\tilde{\boldsymbol{\phi}}^{s+1}))$;
\State $R_{sum}^{\rm ric^{\circ^{s+1}}}=f(\boldsymbol{\phi}^{s+1})$; Update $s=s+1$;
\State \textbf{Until} $||R_{sum}^{\rm ric^{\circ^{s}}}-\bar{R}_{sum}^{\rm ric^{\circ}}||^2< \epsilon$; \textbf{Output} $\boldsymbol{\phi}^*=\boldsymbol{\phi}^{s}$;
\end{algorithmic}
\end{algorithm}

\textit{(P1)} is a constrained maximization problem that can be solved using projected gradient ascent as outlined in Algorithm \ref{alg:euclid}. We increase the objective function by  iteratively updating the RISs phase-shift vector $\boldsymbol{\phi}^{s}$ at iteration $s$ in a step proportional to the positive gradient $\mathbf{p}^{s}$ as $\tilde{\boldsymbol{\phi}}^{s+1} =\boldsymbol{\phi}^{s}+\mu \mathbf{p}^{s}$, where the step size $\mu$ is obtained using backtracking line search. The solution $\tilde{\boldsymbol{\phi}}^{s+1}$ is projected to the closest feasible point satisfying  (\ref{constraint7}) as $\boldsymbol{\phi}^{s+1}=\exp(j \text{arg } (\tilde{\boldsymbol{\phi}}^{s+1}))$ \cite{annie}. Since \textit{(P1)} is a non-convex problem, gradient ascent   only provides a local optimum, but we verify the large gains yielded by  the proposed design in simulations.  The derivative of $R_{sum}^{\rm ric^{\circ}}$, which is the objective in \eqref{P1},  with  $\gamma_k^{\rm ric^{\circ}}$ given by \eqref{det_SINR_RIC} under  MMSE-DFT  protocol,  with respect to $\phi_{ln}$ is  
\begin{align}
&\frac{\partial R_{sum}^{\rm ric^{\circ}}}{\partial \phi_{ln}}=\eta  \sum_{k=1}^{K} \frac{2 d_k \sqrt{\frac{p_{k}q_{k}}{K}}(\text{tr}(\bar{\mathbf{h}}_{0lkn}\bar{\mathbf{h}}_{dk}^{H})+\text{tr}(\bar{\mathbf{h}}_{dk}\bar{\mathbf{h}}_{0lkn}^{H})+2\text{tr}(\sum_{l=1}^L \sum_{i=1}^N \bar{\mathbf{h}}_{0lki} \phi_{li} \bar{\mathbf{h}}_{0lkn}^H))- q_{k}d_k'}{(1+\gamma_{k}^{ric^o})\ln(2) d_k^2}, \nonumber
\end{align} 
where $\eta=\left(1-\frac{S \tau_S}{\tau_C}  \right)$, $q_{k}$ and $d_{k}$ are the numerator and the denominator of \eqref{det_SINR_RIC} respectively, $\bar{\mathbf{h}}_{0lki}=\bar{\mathbf{h}}_{1li}\bar{h}_{2lki}$, $\bar{\mathbf{h}}_{1li}$ is the $i^{th}$ column of $\mathbf{H}_{1l}$, $\bar{h}_{2lki}$ is the $i^{th}$ element of $\bar{\mathbf{h}}_{2lk}$ and
\begin{align}
\begin{split}
d'_k&=\sum_{f\neq k} \frac{p_{f}}{K} \Big(\text{tr} \left( (\mathbf{D}_{k}+\mathbf{A}^{\rm ric}_k)( 2\sum_{l=1}^L \sum_{i=1}^N \phi_{li} \bar{\mathbf{h}}_{0lfi}\bar{\mathbf{h}}_{0lfn}^H+2\bar{\mathbf{h}}_{0lfn} \bar{\mathbf{h}}_{df}^H)\right)+\text{tr}\Big((\mathbf{D}_{f}+\mathbf{C}_{f}^{\rm ric})\\
&\quad \hspace{-.05in}\times(2\sum_{l=1}^L\sum_{i=1}^N \phi_{li}\bar{\mathbf{h}}_{0lki}\bar{\mathbf{h}}_{0lkn}^{H}+\bar{\mathbf{h}}_{0lkn}\bar{\mathbf{h}}_{dk}^{H})\Big) \Big)+\sum_{k=1}^K\frac{p_k}{K \rho}  \text{tr}\left( 2\sum_{l=1}^L\sum_{i=1}^N \phi_{li}\bar{\mathbf{h}}_{0lki}\bar{\mathbf{h}}_{0lkn}^{H}+\bar{\mathbf{h}}_{0lkn}\bar{\mathbf{h}}_{dk}^{H} \right)
\end{split} \nonumber
\end{align} 

 The derivative of $R_{sum}^{\rm ric^{\circ}}$, which is the objective in \eqref{P1},  with  $\gamma_k^{\rm ric^{\circ}}$ given  by \eqref{det_SINR_DE_ric}  under the MMSE-DE CE protocol, with respect to $\phi_{ln}$ is given as
\begin{align}
\label{der_rate_DE}
&\frac{\partial R_{sum}^{\rm ric^{\circ}}}{\partial \phi_{ln}}=\eta\sum_{k=1}^{K} \frac{2 d_k \sqrt{\frac{p_{k}q_{k}}{K}}(2\text{tr}(\sum_{l=1}^L \sum_{i=1}^N\phi_{li}\bar{\mathbf{h}}_{0lki}\bar{\mathbf{h}}_{0lkn}^{H})+ \text{tr}(\bar{\mathbf{h}}_{0lkn}\bar{\mathbf{h}}_{dk}^{H}))- q_{k}d_k'}{(1+\gamma_{k}^{ric^o})\ln(2) d_k^2},
\end{align} 
where $q_{k}$ and $d_{k}$ are the numerator and denominator of \eqref{det_SINR_DE_ric}, respectively, and
\begin{align}
\begin{split}
d'_k&=\sum_{f\neq k} \frac{p_{f}}{K} \Big(\text{tr} \left( (\mathbf{D}_{f}+\mathbf{R}^{\rm ric}_{f} \mathbf{Q}^{\rm ric}_{f}\mathbf{R}^{\rm ric}_{f})(2\sum_{l=1}^L \sum_{i=1}^N \phi_{li}\bar{\mathbf{h}}_{0lki}\bar{\mathbf{h}}_{0lkn}^{H}+2\bar{\mathbf{h}}_{0lkn}\bar{\mathbf{h}}_{dk}^{H})\right)+ \text{tr} \Big( (\mathbf{D}_{k}+\mathbf{R}^{\rm ric}_{k})\\
&\times (2\sum_{l=1}^L\sum_{i=1}^N \phi_{li}\bar{\mathbf{h}}_{0lfi}\bar{\mathbf{h}}_{0lfn}^{H}+\bar{\mathbf{h}}_{0lfn}\bar{\mathbf{h}}_{df}^{H})\Big)\Big)+\sum_{k=1}^K\frac{p_k}{K \rho}  \text{tr}\left(2\sum_{l=1}^L\sum_{i=1}^N \phi_{li}\bar{\mathbf{h}}_{0lki}\bar{\mathbf{h}}_{0lkn}^{H}+\bar{\mathbf{h}}_{0lkn}\bar{\mathbf{h}}_{dk}^{H}\right). \nonumber
\end{split}
\end{align}
The proof of both derivatives follow from direct application of complex derivatives in \cite[4.1]{Pet2008}.

 Using these derivatives along with Algorithm 1, we can optimize reflect beamforming at the RISs to improve the ergodic net sum-rate of the system using only S-CSI.

\subsection{ Instantaneous Net Sum-Rate Maximization Using Full I-CSI}

Since we obtain the full I-CSI of all channels under the MMSE-DFT protocol, we formulate the instantaneous achievable net sum-rate expression and propose to  maximize it using a genetic algorithm in this section, as a performance benchmark  for the S-CSI based RISs design. Recall that the received signal in \eqref{y_k} under MRT precoding can be written as $y_k=\sum_{f=1}^K \zeta \sqrt{p_f} \mathbf{h}_k^{{\rm ric}^H} \hat{\mathbf{h}}^{\rm ric}_f s_f+n_k$. Since the BS does not know the true channels and only has the  estimates, we can write $y_k$ as
\begin{align}
\label{y_kinst1}
&y_k= \zeta \sqrt{p_k} \hat{\mathbf{h}}_k^{{\rm ric}^H} \hat{\mathbf{h}}^{\rm ric}_k s_k+ \zeta \sum_{f\neq k}^K \sqrt{p_f} \hat{\mathbf{h}}_k^{{\rm ric}^H} \hat{\mathbf{h}}^{\rm ric}_f s_f+ \zeta \sum_{f=1}^K \sqrt{p_f} \tilde{\mathbf{h}}_k^{{\rm ric}^H} \hat{\mathbf{h}}^{\rm ric}_f s_f+n_k
\end{align}
where $ \tilde{\mathbf{h}}^{\rm ric}_k$ is the estimation error, that is independent of the MMSE estimate $\hat{\mathbf{h}}^{\rm ric}_k$, and is distributed as $\tilde{\mathbf{h}}_k^{\rm ric}\sim \mathcal{CN}(\mathbf{0}, \tilde{\mathbf{C}}^{\rm ric}_k)$ where $\tilde{\mathbf{C}}_k^{\rm ric}=\mathbf{A}_k^{\rm ric}-\mathbf{C}_k^{\rm ric}$, and $\mathbf{A}_k^{\rm ric}$ and $\mathbf{C}_k^{\rm ric} $ are defined in \eqref{eq_ch1_RIC22} and Lemma \ref{L1_RIC} respectively. Note that  $\hat{\mathbf{h}}^{\rm ric}_k$ and $\tilde{\mathbf{h}}^{\rm ric}_k$ are functions of  $\boldsymbol{\Theta}_l$'s as evident in their expressions in Lemma 2. Treating the last two terms  in \eqref{y_kinst1} as uncorrelated effective noise and  assuming I-CSI to be available at  users, the  instantaneous net rate of user $k$ is presented below.
\begin{theorem}
An achievable instantaneous net rate expression of user $k$ under the MMSE-DFT CE protocol is $R^{\text{inst}}_{k}=\left(1-\frac{S \tau_S}{\tau_C}  \right) \log_2(1+\gamma^{\text{inst}}_{k})$, where the instantaneous SINR  $\gamma^{\text{inst}}_{k}$ is given as
\begin{align}
\label{SINR_MRTinst}
& \gamma^{\text{inst}}_{k}=\frac{p_{k} |\hat{\mathbf{h}}_{k}^{{\rm ric}^H} \hat{\mathbf{h}}^{\rm ric}_{k}|^{2}}{\sum_{f\neq k} p_{f} |\hat{\mathbf{h}}_{k}^{{\rm ric}^H} \hat{\mathbf{h}}^{\rm ric}_{f}|^{2}+ \sum_{f=1}^K p_{f} \hat{\mathbf{h}}_f^{{\rm ric}^H} \tilde{\mathbf{C}}^{\rm ric}_k \hat{\mathbf{h}}^{\rm ric}_f + \frac{\Psi^{\text{inst}}}{\rho}},
\end{align}
where $\Psi^{\text{inst}}={\rm{tr}\left( {\mathbf{P}\hat {\mathbf{H}}^{\rm ric}}{ \hat {\mathbf{H}}^{{\rm ric}^H}} \right)}$ as defined in Sec. II-C and $\rho=\frac{P_{max}}{\sigma^2}$.
\end{theorem}
\begin{IEEEproof}
Treating the last two terms in \eqref{y_kinst1} as noise, we compute $\mathbb{E}[n_kn_k^H]=\sigma^2$, and $\sum_{f=1}^K {p_f} \mathbb{E}_{\tilde{\mathbf{h}}^{\rm ric}_k}[|\tilde{\mathbf{h}}_k^{{\rm ric}^H} \hat{\mathbf{h}}^{\rm ric}_f|^2]=\sum_{f=1}^K {p_f} \mathbb{E}_{\tilde{\mathbf{h}}^{\rm ric}_k}[\tilde{\mathbf{h}}_k^{{\rm ric}^H} \hat{\mathbf{h}}^{\rm ric}_f \hat{\mathbf{h}}_f^{{\rm ric}^H} \tilde{\mathbf{h}}^{\rm ric}_k]=\sum_{f=1}^K {p_f} tr(\tilde{\mathbf{C}}^{\rm ric}_k\hat{\mathbf{h}}^{\rm ric}_f \hat{\mathbf{h}}_f^{{\rm ric}^H})$.
\end{IEEEproof}
The instantaneous net sum-rate is then given as  \vspace{-.05in}
\begin{align}
\label{rate_suminst}
R^{\text{inst}}_{sum}=\sum_{k=1}^K \left(1-\frac{S \tau_S}{\tau_C}  \right) \log_2(1+\gamma^{\text{inst}}_{k}),
\end{align}


Next, we formulate the   net sum-rate maximization problem similar to Problem \textit{(P1)} with the objective $\sum_{k=1}^K \left(1-\frac{S \tau_S}{\tau_C}  \right) \log_2(1+\gamma^{\text{inst}}_{k})$. The RIS phase shifts are designed to solve this problem using the genetic algorithm outlined in \cite{GA}. Note that this full I-CSI based RISs design is expected to yield high sum-rate as the phase shifts are designed for each channel realization to meet the desired objective. On the other hand, S-CSI based RISs design  only optimizes the phase shifts when channel statistics change. However the I-CSI based RISs design requires full CSI of all channels, i.e. $\hat{\mathbf{h}}_{dk}$ and $\hat{\mathbf{h}}_{2lk}$'s, which imposes a large training overhead of $S=NL/M+1$ sub-phases. This would compromise the net sum-rate, as we will see in the simulations next, because the training loss factor represented by $\left(1-\frac{S \tau_S}{\tau_C}  \right)$ in \eqref{rate_suminst} will reduce as $N$ increases. 


\vspace{-.14in}
\section{Simulation Results}
To generate the simulation results, the  BS equipped with an $M$-antenna ULA is considered to be deployed at $(0,0,0)$m along the z-axis, where $(x,y,z)$ denotes the Cartesian coordinates. The $K$ users are placed along an arc of radius $400$m that spans angles from $-30^{\circ}$ to $30^{\circ}$ with respect to the $y$-axis,  and the multiple RISs are placed equidistantly on an arc of radius $250$m while  spanning angles from $-30^{\circ}$ to $30^{\circ}$ with respect to the $y$-axis. Each RIS planar array is placed in the x-z plane.  We define $p_k=1/K$, $\forall k$ and $P_{max}=10$ W. The path loss model is  represented as $\beta_{k}=\frac{C_0}{d^{\bar{\alpha}}}$, with $C_0$ set as $30$\rm{dB} for all links. The  receiver noise variance $\sigma^2$ is $-94$dBm. The path loss coefficient for the BS-RIS LoS links is set as $\bar{\alpha}_{1l}=2$, that  for RIS-user links is set as  $\bar{\alpha}_{2lk}=2.8$, and that for BS-user links is set as $\bar{\alpha}_{dk}=3.5$ \cite{Wu1}. The coherence interval is set as $\tau_c=2000$ symbols. The Rician factor is calculated as $\kappa_{ek}= 13-0.03d_{ek}$, where $e \in \{d, l\}$ and $d_{ek}$ is the associated link distance. The rest of the parameters are mentioned under each figure. 

\vspace{-.1in}
\subsection {Simulation Results under Rician Fading}

\renewcommand{\baselinestretch}{1.5}

\begin{figure*}[!t]
\begin{minipage}[b]{0.48\linewidth}
\centering
\tikzset{every picture/.style={scale=.95}, every node/.style={scale=.8}}
%
%
\definecolor{mycolor1}{rgb}{0.60000,0.20000,0.00000}%
\definecolor{mycolor2}{rgb}{0.00000,0.49804,0.00000}%
\begin{tikzpicture}

\begin{axis}[%
width=.85\textwidth,
height=.75\textwidth,
scale only axis,
xmin=1,
xmax=20,
xlabel style={font=\color{white!15!black}},
xlabel={$P_{max}$ $(W)$},
ymin=0,
ymax=0.14,
ylabel style={font=\color{white!15!black}},
ylabel={Average SINR},
axis background/.style={fill=white},
xmajorgrids,
ymajorgrids,
legend style={at={(axis cs: 1,0.14)},anchor=north west,legend cell align=left,align=left,draw=white!15!black, /tikz/column 2/.style={
                column sep=5pt,
            }},]
\addplot [color=mycolor1, line width=1.0pt,  mark=triangle, mark options={solid, rotate=270, mycolor1}]
  table[row sep=crcr]{%
2	0.00856584687550789\\
4	0.0247966910617771\\
6	0.0399482335224199\\
8	0.0541449002705827\\
10	0.0674910710785844\\
12	0.0800751817897544\\
14	0.0919728216011166\\
16	0.10324910843985\\
18	0.113960536410369\\
20	0.124156430881901\\
};
\addlegendentry{\footnotesize Perfect CSI-opt. RISs (Th)}

\addplot [color=mycolor1, line width=1.0pt,  mark=square, mark size=2.5pt, mark options={solid, mycolor1}]
  table[row sep=crcr]{%
2	0.00736995835550143\\
4	0.0211261391739996\\
6	0.0337317002481608\\
8	0.0453457070766677\\
10	0.0560970642256904\\
12	0.0660916867722567\\
14	0.0754176512788026\\
16	0.084148975312096\\
18	0.0923484431635327\\
20	0.100069753836228\\
};
\addlegendentry{\footnotesize  DFT-CE-opt. RISs (Th)}

\addplot [color=mycolor1, line width=1.0pt, mark=o,mark size=2.5pt, mark options={solid, mycolor1}]
  table[row sep=crcr]{%
2	0.0071776514775642\\
4	0.0205270809218494\\
6	0.0327066176231697\\
8	0.043884135899965\\
10	0.0541945969128743\\
12	0.0637481180606846\\
14	0.0726357034002355\\
16	0.0809334066497414\\
18	0.0887054163192941\\
20	0.0960063825449584\\
};
\addlegendentry{\footnotesize DE-CE-opt. RISs (Th)}

\addplot [color=mycolor2, dashed, line width=1.0pt,  mark=x, mark options={solid, mycolor2}]
  table[row sep=crcr]{%
2	0.00742223256720182\\
4	0.0211448283690985\\
6	0.0338130073467997\\
8	0.0454201604684503\\
10	0.0563967193607011\\
12	0.0660555132921863\\
14	0.0755187049507112\\
16	0.0845732892445153\\
18	0.0930317621556364\\
20	0.10072866213704\\
};
\addlegendentry{\footnotesize DFT-CE-opt. RISs (MC)}

\addplot [color=mycolor2, dashed, line width=1.0pt, mark=+, mark options={solid, mycolor2}]
  table[row sep=crcr]{%
2	0.00722128578025714\\
4	0.0205325538026032\\
6	0.0327428081972574\\
8	0.0439068692945081\\
10	0.0544161878206971\\
12	0.063628684321747\\
14	0.0726875766639627\\
16	0.0813461419611141\\
18	0.0893031917225323\\
20	0.0966492186452109\\
};
\addlegendentry{\footnotesize DE-CE-opt. RISs (MC)}

\addplot [color=red, line width=1.0pt,  mark=triangle, mark options={solid, rotate=270, red}]
  table[row sep=crcr]{%
2	0.00499781549553358\\
4	0.0149377593075945\\
6	0.0248041923234659\\
8	0.0345979877612694\\
10	0.0443200039404214\\
12	0.0539710846232396\\
14	0.0635520593465378\\
16	0.0730637437435694\\
18	0.0825069398566627\\
20	0.0918824364408801\\
};
\addlegendentry{\footnotesize Perfect CSI-rand. RISs (Th)}

\addplot [color=red, line width=1.0pt,  mark=square, mark options={solid, red}]
  table[row sep=crcr]{%
2	0.00347844758878455\\
4	0.0103949999512915\\
6	0.0172583530998494\\
8	0.0240692040833591\\
10	0.0308282361343473\\
12	0.0375361190486481\\
14	0.0441935095516164\\
16	0.0508010516514571\\
18	0.057359376980225\\
20	0.0638691051230199\\
};
\addlegendentry{\footnotesize DFT-CE-rand. RISs (Th)}

\addplot [color=red, line width=1.0pt,  mark=o, mark options={solid, red}]
  table[row sep=crcr]{%
2	0.00321876152411253\\
4	0.00961836798302375\\
6	0.0159679933521135\\
8	0.0222683142520297\\
10	0.0285199932652871\\
12	0.0347236793422093\\
14	0.0408800081917246\\
16	0.0469896026577022\\
18	0.0530530730814829\\
20	0.0590710176512216\\
};
\addlegendentry{\footnotesize DE-CE-rand. RISs (Th)}

\end{axis}

\end{tikzpicture}%
\caption{Validation of  deterministic equivalents of  SINR for  $M,N=60$ and $L,K=20$. } 
\label{Fig2a}
\end{minipage}
\hspace{.4cm}
\begin{minipage}[b]{0.48\linewidth}
\centering
\tikzset{every picture/.style={scale=.95}, every node/.style={scale=.8}}
%
%
\definecolor{mycolor1}{rgb}{0.60000,0.20000,0.00000}%
\definecolor{mycolor2}{rgb}{0.00000,0.49804,0.00000}%
\begin{tikzpicture}

\begin{axis}[%
width=.85\textwidth,
height=.75\textwidth,
scale only axis,
xmin=1,
xmax=20,
xlabel style={font=\color{white!15!black}},
xlabel={$P_{max}$ $(W)$},
ymin=0,
ymax=4,
ylabel style={font=\color{white!15!black}},
ylabel={Net sum rate (bps/Hz)},
axis background/.style={fill=white},
xmajorgrids,
ymajorgrids,
legend style={at={(axis cs:1,4)},anchor=north west,legend cell align=left,align=left,draw=white!15!black, /tikz/column 2/.style={
                column sep=5pt,
            }},]
            
            \addplot [color=mycolor1, line width=1.0pt, mark=triangle, mark options={solid, rotate=270, mycolor1}]
  table[row sep=crcr]{%
2	0.245478208632131\\
4	0.701899896033796\\
6	1.1185230499045\\
8	1.50135698780854\\
10	1.85514757649617\\
12	2.18371224736982\\
14	2.490169435789\\
16	2.77710045158634\\
18	3.04666658486738\\
20	3.300695671252\\
};
\addlegendentry{\footnotesize Perfect CSI-opt. RISs (Th)}

\addplot [color=mycolor1,  line width=1.0pt,  mark=square, mark size=2.5pt, mark options={solid, mycolor1}]
  table[row sep=crcr]{%
2	0.175721587390784\\
4	0.497627734797997\\
6	0.786253233815036\\
8	1.04729449094452\\
10	1.28511686789276\\
12	1.50314205054182\\
14	1.70410272579035\\
16	1.89021607839448\\
18	2.06330554766472\\
20	2.22488845421191\\
};
\addlegendentry{\footnotesize DFT-CE-opt. RISs (Th)}

\addplot [color=mycolor1,  line width=1.0pt,  mark=o, mark size=2.5pt,mark options={solid, mycolor1}]
  table[row sep=crcr]{%
2	0.204042387327688\\
4	0.576491540191524\\
6	0.909003960063153\\
8	1.20860290261588\\
10	1.48063220968529\\
12	1.72925513524674\\
14	1.95777914879811\\
16	2.16887539907998\\
18	2.36473164424908\\
20	2.54716163551654\\
};
\addlegendentry{\footnotesize DE-CE-opt. RISs (Th)}

\addplot [color=mycolor2, dashed, line width=1.0pt,  mark=x, mark options={solid, mycolor2}]
  table[row sep=crcr]{%
2	0.176952760147692\\
4	0.498059736378242\\
6	0.788019621992002\\
8	1.04898945950002\\
10	1.29199466294081\\
12	1.50228410867052\\
14	1.70632512852647\\
16	1.89940523762325\\
18	2.07744049017999\\
20	2.23847212952827\\
};
\addlegendentry{\footnotesize DFT-CE-opt. RISs (MC)}

\addplot [color=mycolor2, dashed, line width=1.0pt,  mark=+, mark options={solid, mycolor2}]
  table[row sep=crcr]{%
2	0.205267347929061\\
4	0.576644996483976\\
6	0.909959098899138\\
8	1.2092175538174\\
10	1.4867874897742\\
12	1.7262227729363\\
14	1.95937481317\\
16	2.17977865466275\\
18	2.37945338523826\\
20	2.56348693622406\\
};
\addlegendentry{\footnotesize DE-CE-opt. RISs (MC)}

\addplot [color=red, line width=1.0pt,  mark=triangle, mark options={solid, rotate=270, red}]
  table[row sep=crcr]{%
2	0.143838287894175\\
4	0.427746431543342\\
6	0.706753572510684\\
8	0.981002421525021\\
10	1.25062964427261\\
12	1.51576620290584\\
14	1.77653767322118\\
16	2.03306453960971\\
18	2.28546246966876\\
20	2.53384257017448\\
};
\addlegendentry{\footnotesize Perfect CSI-rand. RISs (Th)}

\addplot [color=red,  line width=1.0pt,  mark=square, mark options={solid, red}]
  table[row sep=crcr]{%
2	0.0833524256238262\\
4	0.248190048549565\\
6	0.410590430215234\\
8	0.570614396741367\\
10	0.728320538674703\\
12	0.883765322668987\\
14	1.03700319598229\\
16	1.18808668436209\\
18	1.33706648383512\\
20	1.48399154687033\\
};
\addlegendentry{\footnotesize DFT-CE-rand. RISs (Th)}

\addplot [color=red,  line width=1.0pt,  mark=o, mark options={solid, red}]
  table[row sep=crcr]{%
2	0.0919734719746468\\
4	0.273908939117882\\
6	0.453217760851156\\
8	0.62996419215052\\
10	0.804210160185253\\
12	0.976015379642115\\
14	1.14543746065381\\
16	1.31253200992026\\
18	1.47735272555524\\
20	1.63995148614071\\
};
\addlegendentry{\footnotesize DE-CE-rand. RISs (Th)}

\addplot [color=black, line width=1.0pt, mark=star, mark options={solid, black}]
  table[row sep=crcr]{%
2	0.101902158938797\\
4	0.303954237256323\\
6	0.503708216559567\\
8	0.701208816245816\\
10	0.896499491534371\\
12	1.08962248075902\\
14	1.28061885045159\\
16	1.46952853834049\\
18	1.65639039437996\\
20	1.84124221991832\\
};
\addlegendentry{\footnotesize No RISs-Perfect CSI}

\addplot [color=black, line width=1.0pt,  mark=diamond, mark options={solid, black}]
  table[row sep=crcr]{%
2	0.0565316385784911\\
4	0.168882780122322\\
6	0.280296635009907\\
8	0.390787570873041\\
10	0.500369616659857\\
12	0.609056473639134\\
14	0.716861525945939\\
16	0.823797850691992\\
18	0.929878227662664\\
20	1.03511514862125\\
};
\addlegendentry{\footnotesize No RISs-Imperfect CSI}

\end{axis}
\end{tikzpicture}%
\caption{Validation of  deterministic equivalents of  net sum-rate for \hspace{-.02in}$M,N=60$ and \hspace{-.01in}$L,K=20$.}
\label{Fig2b}
\end{minipage}
\end{figure*}

We  first validate the tightness of the deterministic equivalents of the SINR under Rician fading  in Fig. \ref{Fig2a}.  The theoretical (Th) deterministic equivalents of the SINR under MMSE-DFT CE  in Theorem \ref{thm1_ric}, under MMSE-DE CE  in Theorem \ref{thm2_ric}, and under perfect CSI  in Corollary \ref{cor_perric} are plotted under random (rand.) phase shifts at all RISs as well as under RISs phase shifts optimized (opt.) using Algorithm 1 based on S-CSI.      We also plot on this figure the Monte-Carlo (MC) simulated SINR values in \eqref{SINR_MRT} under  both CE protocols for I-CSI acquisition that is used to implement precoding, while considering  Algorithm 1 to design RISs phase shifts using S-CSI. The figure shows an excellent match between the Monte-Carlo simulated SINR values and its deterministic equivalents under both CE protocols, even for moderate system sizes of $M=60$ and $K=20$. Therefore the deterministic equivalents are a powerful tool for the performance analysis of massive MIMO systems, without relying on time-consuming Monte-Carlo simulations. 

Next we observe that optimizing the RISs phase shifts using Algorithm 1 based only on S-CSI yields significant SINR gains over choosing these phase shifts randomly.  We  also observe  that the SINR values are higher under MMSE-DFT CE protocol than those under MMSE-DE protocol, where both these protocols are used to construct the estimate of the aggregate channel $\mathbf{h}^{\rm ric}_k$ for MRT precoding. This is because the former achieves a much better estimation quality by using an optimal DFT based solution for the phase shifts at all RISs during the CE phase. 



Note that Fig. 2 does not capture the penalty due to the channel training overhead since it only plots the average SINR. To take this into account,  in Fig. \ref{Fig2b}, we plot the ergodic net sum-rate in (\ref{R_sum_ric}) using the deterministic equivalent of the SINR in Theorem \ref{thm1_ric} for MMSE-DFT, the one in Theorem \ref{thm2_ric} for MMSE-DE, and the one in Corollary \ref{cor_perric} for perfect CSI, under random and optimized (Alg. 1) RISs phase shifts. We also plot the Monte-Carlo simulated net sum-rate in (\ref{R_sum_MC}) under  both CE protocols using optimized (Alg. 1) RISs phase shifts. In contrast to the observation from Fig. \ref{Fig2a}, we see here  that the net sum-rate under MMSE-DFT is lower than that under MMSE-DE CE protocol, where both protocols are used for I-CSI acquisition to implement precoding. This is because  the higher SINR obtained under the MMSE-DFT CE protocol due to the better channel estimation quality comes at the expense of a large training overhead of $S\tau_S =(NL/M +1)K$ symbols to construct $\hat{\mathbf{h}}^{\rm ric}_k$'s using full CSI. On the other hand, the MMSE-DE protocol estimates all $\mathbf{h}^{\rm ric}_k$'s in just $\tau_S=K$ symbols. The impact of larger training overhead on  the net sum-rate in (\ref{R_sum_ric}) compromises the effect of the increased  SINR under  MMSE-DFT CE protocol, resulting in lower net sum-rate values as compared to those under DE  protocol.

Comparing the net sum-rate values under the optimized RISs-assisted system and the system without RISs in Fig. \ref{Fig2b}, we see that using RISs introduces a  $3.6\times$ gain in the net sum-rate for $P_{max}=2$W, while this gain is around $2.4\times$ for $P_{max}=20$W. Therefore the proposed RISs phase shifts design based on only S-CSI yields significant performance gains in both noise and interference-limited scenarios under Rician fading compared to a system without RISs, making a RISs a promising energy efficient solution for future mobile broadband networks. 

\begin{figure*}[!t]
\begin{minipage}[b]{0.48\linewidth}
\centering
\tikzset{every picture/.style={scale=.95}, every node/.style={scale=.8}}
%
%
\definecolor{mycolor1}{rgb}{0.60000,0.20000,0.00000}%
\definecolor{mycolor2}{rgb}{1.00000,0.00000,1.00000}%
\begin{tikzpicture}

\begin{axis}[%
width=.9\textwidth,
height=.8\textwidth,
scale only axis,
xmin=0,
xmax=350,
xlabel style={font=\color{white!15!black}},
xlabel={$N$},
ymin=0.2,
ymax=1.4,
ylabel style={font=\color{white!15!black}},
ylabel={Net sum rate (bps/Hz)},
axis background/.style={fill=white},
xmajorgrids,
ymajorgrids,
legend style={at={(axis cs: 0,1.4)},anchor=north west,legend cell align=left,align=left,draw=white!15!black, /tikz/column 2/.style={
                column sep=5pt,
            }},]
\addplot [color=mycolor2, line width=1.0pt,mark=x, mark options={solid, mycolor2}]
  table[row sep=crcr]{%
20	0.349449235725966\\
60	0.570619686233413\\
80	0.661287749054878\\
100	0.690142889208297\\
160	0.630403836272409\\
240	0.572933374552736\\
320	0.507651962749354\\
};
\addlegendentry{\footnotesize DFT-CE (MRT), I-CSI (RISs)}

\addplot [color=mycolor1, line width=1.0pt,  mark=square, mark options={solid, mycolor1}]
  table[row sep=crcr]{%
20	0.314075851834456\\
60	0.508389231895292\\
80	0.582193661130019\\
100	0.606773646241047\\
160	0.571757693727538\\
240	0.52285484818227\\
320	0.476386131211849\\
};
\addlegendentry{\footnotesize DFT-CE (MRT), S-CSI (RISs)}

\addplot [color=mycolor1, line width=1.0pt, mark=o, mark options={solid, mycolor1}]
  table[row sep=crcr]{%
20	0.315403864875432\\
60	0.534703991773895\\
80	0.627650086401156\\
100	0.669878688293443\\
160	0.742971246432717\\
240	0.784401288512283\\
320	0.820857547318518\\
};
\addlegendentry{\footnotesize DE-CE (MRT), S-CSI (RISs)}

\addplot [color=mycolor1, line width=1.0pt, mark=triangle, mark options={solid, rotate=270, mycolor1}]
  table[row sep=crcr]{%
20	0.542652905048024\\
60	0.757955204618245\\
80	0.852926038367936\\
100	0.897578573839849\\
160	0.977993223514919\\
240	1.0205452093156334\\
320	1.08002031426666\\
};
\addlegendentry{\scriptsize Perfect CSI (MRT), S-CSI (RISs)}

\addplot [color=black, dashed, line width=1.0pt, mark=o, mark options={solid, black}]
  table[row sep=crcr]{%
20	0.280296635009907\\
60	0.280296635009907\\
80	0.280296635009907\\
100	0.280296635009907\\
160	0.280296635009907\\
240	0.280296635009907\\
320	0.280296635009907\\
};
\addlegendentry{\footnotesize No RISs, imperfect CSI}

\addplot [color=black, dashed, line width=1.0pt, mark=triangle, mark options={solid, rotate=270, black}]
  table[row sep=crcr]{%
20	0.503708216559567\\
60	0.503708216559567\\
80	0.503708216559567\\
100	0.503708216559567\\
160	0.503708216559567\\
240	0.503708216559567\\
320	0.503708216559567\\
};
\addlegendentry{\footnotesize No RISs, perfect CSI}

\end{axis}

\begin{axis}[
width=.9\textwidth,
height=.8\textwidth,
  axis y line*=right,
  axis x line=none,
	scale only axis,
  ymin=0, ymax=2200,
  ylabel=Training overhead (symbols),
  ylabel style={font=\color{blue}},
    legend style={at={(axis cs: 350,0)},anchor=south east,legend cell align=left,align=left,draw=white!15!black, /tikz/column 2/.style={
                column sep=5pt,
            }},]
]
\addplot [color=blue, line width=1.0pt, mark=pentagon, mark options={solid, blue}]
  table[row sep=crcr]{%
20	153\\
60	420\\
80	553\\
100	687\\
160	1086\\
240	1620\\
320	2153\\
};
\addlegendentry{\footnotesize  DFT-CE}
420        1220        1620        2020        3220        4820        6420
\addplot [color=blue, line width=1.0pt, mark=diamond, mark options={solid, blue}]
  table[row sep=crcr]{%
20	20\\
60	20\\
80	20\\
100	20\\
160	20\\
240	20\\
320	20\\
};
\addlegendentry{\footnotesize  DE-CE}

\end{axis}
\end{tikzpicture}%
\caption{Net sum-rate and training overhead  for $M=60$, $K=20$ and $L=20$. } 
\label{Fig5}
\end{minipage}
\hspace{.8cm}
\begin{minipage}[b]{0.48\linewidth}
\centering
\tikzset{every picture/.style={scale=.95}, every node/.style={scale=.8}}
%
%
\definecolor{mycolor1}{rgb}{0.60000,0.20000,0.00000}%
\definecolor{mycolor2}{rgb}{0.00000,0.49804,0.00000}%
\begin{tikzpicture}

\begin{axis}[%
width=.87\textwidth,
height=.8\textwidth,
scale only axis,
xmin=0,
xmax=40,
xlabel style={font=\color{white!15!black}},
xlabel={Number of RISs $L$},
ymin=0.6,
ymax=2.7,
ylabel style={font=\color{white!15!black}},
ylabel={Net sum rate (bps/Hz)},
axis background/.style={fill=white},
xmajorgrids,
ymajorgrids,
legend style={at={(axis cs: 0,2.7)},anchor=north west,legend cell align=left,align=left,draw=white!15!black, /tikz/column 2/.style={
                column sep=5pt,
            }},]
\addplot [color=mycolor1, line width=1.0pt,  mark=square, mark options={solid, mycolor1}]
  table[row sep=crcr]{%
2	0.670655270335422\\
4	0.915211375748191\\
10	1.10982760990205\\
20	1.31341895683944\\
30	1.27117416825761\\
40	1.21592513670046\\
};
\addlegendentry{\footnotesize Dist. RISs, DFT-CE}

\addplot [color=mycolor1, line width=1.0pt,  mark=o, mark options={solid, mycolor1}]
  table[row sep=crcr]{%
2	0.72966837392514\\
4	1.02292356214467\\
10	1.25793833073007\\
20	1.50825339853079\\
30	1.45571656995307\\
40	1.38651329096449\\
};
\addlegendentry{\footnotesize Dist. RISs, DE-CE}

\addplot [color=mycolor1, line width=1.0pt, mark=triangle, mark options={solid, rotate=270, mycolor1}]
  table[row sep=crcr]{%
2	1.23755713749576\\
4	1.50826000191877\\
10	1.71962001737583\\
20	1.92692336186172\\
30	1.88561740591633\\
40	1.83426097322467\\
};
\addlegendentry{\footnotesize Dist. RISs, perfect CSI}

\addplot [color=mycolor2, dashed, line width=1.0pt,  mark=square, mark options={solid, mycolor2}]
  table[row sep=crcr]{%
2	0.626509077105397\\
4	0.626509077105397\\
10	0.626509077105397\\
20	0.626509077105397\\
30	0.626509077105397\\
40	0.626509077105397\\
};
\addlegendentry{\footnotesize Cent. RIS, DFT-CE}

\addplot [color=mycolor2, dashed, line width=1.0pt, mark=o, mark options={solid, mycolor2}]
  table[row sep=crcr]{%
2	0.677451596988237\\
4	0.677451596988237\\
10	0.677451596988237\\
20	0.677451596988237\\
30	0.677451596988237\\
40	0.677451596988237\\
};
\addlegendentry{\footnotesize Cent. RIS, DE-CE}

\addplot [color=mycolor2, dashed, line width=1.0pt, mark=triangle, mark options={solid, rotate=270, mycolor2}]
  table[row sep=crcr]{%
2	1.18312474525913\\
4	1.18312474525913\\
10	1.18312474525913\\
20	1.18312474525913\\
30	1.18312474525913\\
40	1.18312474525913\\
};
\addlegendentry{\footnotesize Cent. RIS, perfect CSI}

\end{axis}
\end{tikzpicture}%
\caption{Net sum-rate against the number of RISs for $M=60$, $K=20$ and $NL=1200$.}
\label{Fig4}
\end{minipage}
\end{figure*}


Next in Fig. \ref{Fig5} we compare the net sum-rate performance of an RISs-assisted system against $N$ for three scenarios: (i) the RISs phase shifts are designed using full I-CSI of all links under MMSE-DFT CE protocol as outlined in Sec. IV-B, (ii) the RISs phase shifts are designed based on S-CSI using Algorithm 1 while considering I-CSI obtained using the MMSE-DFT protocol to implement MRT, and (iii) the RISs phase shifts are designed based on S-CSI using Algorithm 1 while considering I-CSI obtained using MMSE-DE protocol to implement MRT. Note that for scenario (i) the average instantaneous net sum-rate in \eqref{rate_suminst} is plotted, while for scenarios (ii) and (iii) the deterministic equivalents of the ergodic net sum-rate in \eqref{R_sum_ric} are plotted.  Additionally, the ergodic net sum-rate under perfect CSI in Corollary 4, and without RISs in Theorem 3 are also plotted.  The net sum-rate of RISs-assisted system under perfect CSI and under DE CE protocol  increases considerably with $N$, with the gap from the ``No RISs" scenario  becoming significant. Next we observe that the net sum-rate under MMSE-DFT CE protocol increases until a certain point ($N=100$) and then starts to decrease. This is because after this point the increase in the training overhead given by $S\tau_S =(NL/M +1)K$ symbols, plotted in blue on the right y-axis, becomes dominant over the increase in SINR that comes with the use of more reflecting elements, and overall the net sum-rate in \eqref{R_sum_ric} starts to deteriorate. On the other hand the training overhead of the MMSE-DE CE protocol is $\tau_S=K$ symbols as plotted in blue on the figure, irrespective of the values of $N$ and $L$, since the number of required sub-phases $S$ is one. This results in  scenario (iii) to perform better than scenario (ii) due to the significantly improved training loss factor $\left(1-\frac{S\tau_S}{\tau_C}\right)$ in the net sum-rate expression in \eqref{R_sum_ric}.

Next we compare the performance of S-CSI and I-CSI based designs for RISs phase shifts in Sec. IV.  We observe that (i) shows a similar trend as (ii) with the net sum-rate first increasing and then decreasing with $N$ due to the large training overhead incurred by the MMSE-DFT CE protocol. However the net sum-rate is higher under (i) than that under (ii) because we are using full I-CSI to optimize the RISs phase shifts  to realize favorable instantaneous channels that maximize the average instantaneous net sum-rate, instead of optimizing the phase shifts only to realize favorable channel statistics after several coherence periods as done in scenario (ii).  However when comparing all three, scenario (iii) performs the best due to the very low training overhead of the MMSE-DE CE protocol in which the aggregate channel $\mathbf{h}_k^{\rm ric}$ is directly estimated. Even though this scenario will have lower SINR values compared to (i) since we only use S-CSI to design the RISs phase shifts, but overall the MMSE-DE+S-CSI based RISs design  outperforms the I-CSI based RISs design for $N>100$, making it a desirable scheme. 



In Fig. \ref{Fig4} we again plot the deterministic equivalents of the ergodic  net sum-rate under perfect I-CSI, imperfect I-CSI using MMSE-DFT CE protocol, and imperfect I-CSI using MMSE-DE protocol, to implement precoding, while using Algorithm 1 to design the RISs phase shifts based on S-CSI. The curves are plotted against the number of RISs $L$ while keeping the total number of reflecting elements in the whole system constant at $NL=1200$. We also plot on this figure the net sum-rate under the centralized RIS scenario, where all the $NL$ reflecting elements are placed in one central RIS positioned at the center of the arc.  The net sum-rate first increases  with $L$, because as the $NL$ reflecting elements are distributed over the arc, the RISs are able to better exploit the spatial degrees of freedom to improve the performance of the users  who are also distributed and can now have some RISs that are closer than the centralized RIS. However as we divide the $NL$ elements into more than $L=20$ RISs, the net sum-rate starts to drop  under the distributed RISs setup. This is because increasing $L$  beyond $20$  reduces the number of reflecting elements per RIS to less than $60$, which causes  the gain from each RIS to decrease. This decrease in performance gain per RIS starts to dominate over the increase in performance that comes with distributing the RIS elements into more surfaces over a geographical area. As a result for a fixed number of total RIS elements, it is important to balance the elements across the distributed RISs to achieve the optimal performance. However we note that even with $L=40$ RISs each having  $30$ elements, the performance of  distributed RISs setup is  better than  the centralized RIS setup, due to each user having stronger reflected channels with atleast some RISs that are closer to it.

\vspace{-.1in}
\subsection{Simulation Results under Rayleigh Fading}

\begin{figure*}[!t]
\begin{minipage}[b]{0.48\linewidth}
\centering
\tikzset{every picture/.style={scale=.95}, every node/.style={scale=.8}}
%
%
\definecolor{mycolor1}{rgb}{0.60000,0.20000,0.00000}%
\definecolor{mycolor2}{rgb}{0.00000,0.49804,0.00000}%
\begin{tikzpicture}

\begin{axis}[%
width=.9\textwidth,
height=.5\textwidth,
scale only axis,
xmin=0,
xmax=350,
xlabel style={font=\color{white!15!black}},
xlabel={$N$},
ymin=0,
ymax=0.13,
ylabel style={font=\color{white!15!black}},
ylabel={Net sum rate (bps/Hz)},
axis background/.style={fill=white},
scaled ticks = false,
tick label style={/pgf/number format/fixed},
xmajorgrids,
ymajorgrids,
legend style={at={(axis cs: 350,0.13)},anchor=north east,legend cell align=left,align=left,draw=white!15!black, /tikz/column 2/.style={
                column sep=5pt,
            }},]
\addplot [color=mycolor1, line width=1.0pt, mark=+, mark options={solid, mycolor1}]
  table[row sep=crcr]{%
20	0.0293644815838886\\
60	0.0682514183093984\\
80	0.0786270343594136\\
100	0.0869606976136521\\
120	0.0918350963423347\\
160	0.0933789662958618\\
200	0.0871364500367113\\
240	0.0736545040552618\\
280	0.056165750459446\\
320	0.0347958010743063\\
};
\addlegendentry{\scriptsize DFT-CE (Th)}

\addplot [color=mycolor1, line width=1.0pt, mark=x, mark size=2pt,  mark options={solid, mycolor1}]
  table[row sep=crcr]{%
20	0.0051984170176458\\
60	0.00643494649008173\\
80	0.00715264812029986\\
100	0.00793493738199007\\
120	0.00878181722818373\\
160	0.01067216803607\\
200	0.012837237722101\\
240	0.0152871994662953\\
280	0.018023263309681\\
320	0.0210617090601847\\
};
\addlegendentry{\scriptsize DE-CE (Th)}

\addplot [color=red, line width=1.0pt,  mark=square, mark size=2.5pt, mark options={solid, red}]
  table[row sep=crcr]{%
20	0.0293644815838886\\
60	0.0682514183093984\\
80	0.0786270343594136\\
100	0.0869606976136521\\
120	0.0918350963423347\\
160	0.0933789662958618\\
200	0.0871364500367113\\
240	0.0736545040552618\\
280	0.056165750459446\\
320	0.0347958010743063\\
};
\addlegendentry{\scriptsize DFT-CE, rand. RISs (MC)}

\addplot [color=red, line width=1.0pt, mark=o, mark size=2.75pt, mark options={solid, red}]
  table[row sep=crcr]{%
20	0.0051984170176458\\
60	0.00643494649008173\\
80	0.00715264812029986\\
100	0.00793493738199007\\
120	0.00878181722818373\\
160	0.01067216803607\\
200	0.012837237722101\\
240	0.0152871994662953\\
280	0.018023263309681\\
320	0.0210617090601847\\
};
\addlegendentry{\scriptsize DE-CE, rand. RISs (MC)}

\addplot [color=blue, line width=1.0pt, dashdotted]
  table[row sep=crcr]{%
20	0.0293644815838886\\
60	0.0682514183093984\\
80	0.0786270343594136\\
100	0.0869606976136521\\
120	0.0918350963423347\\
160	0.0933789662958618\\
200	0.0871364500367113\\
240	0.0736545040552618\\
280	0.056165750459446\\
320	0.0347958010743063\\
};
\addlegendentry{\scriptsize DFT-CE, opt. RISs (MC)}

\addplot [color=black, line width=1.0pt,dashdotted]
  table[row sep=crcr]{%
20	0.0051984170176458\\
60	0.00643494649008173\\
80	0.00715264812029986\\
100	0.00793493738199007\\
120	0.00878181722818373\\
160	0.01067216803607\\
200	0.012837237722101\\
240	0.0152871994662953\\
280	0.018023263309681\\
320	0.0210617090601847\\
};
\addlegendentry{\scriptsize DE-CE, opt. RISs  (MC)}



\addplot [color=mycolor2,  line width=1.0pt,  mark=diamond, mark options={solid, mycolor2}]
  table[row sep=crcr]{%
20	0.00468493152321193\\
60	0.00468493152321193\\
80	0.00468493152321193\\
100	0.00468493152321193\\
120	0.00468493152321193\\
160	0.00468493152321193\\
200	0.00468493152321193\\
240	0.00468493152321193\\
280	0.00468493152321193\\
320	0.00468493152321193\\
};


\node at (axis cs: 40,.04) [anchor=west] {\footnotesize Green line: No RISs (imperfect CSI)};

\end{axis}
\end{tikzpicture}%
\caption{Net sum-rate  against $N$ under Rayleigh fading for $M=70$, $K=20$, and $L=20$. } 
\label{Fig6}
\end{minipage}
\hspace{.4cm}
\begin{minipage}[b]{0.48\linewidth}
\centering
\tikzset{every picture/.style={scale=.95}, every node/.style={scale=.8}}
%
%
\definecolor{mycolor1}{rgb}{0.00000,0.49804,0.00000}%
\begin{tikzpicture}

\begin{axis}[%
width=.9\textwidth,
height=.5\textwidth,
scale only axis,
xmin=-20,
xmax=20,
xlabel style={font=\color{white!15!black}},
xlabel={$\rho\text{ (dB)}$},
ymin=-10,
ymax=5.5,
ylabel style={font=\color{white!15!black}},
ylabel={Average SINR (dB)},
axis background/.style={fill=white},
xmajorgrids,
ymajorgrids,
legend style={at={(axis cs: 20, -10)},anchor=south east,legend cell align=left,align=left,draw=white!15!black, /tikz/column 2/.style={
                column sep=5pt,
            }},]
						
\addplot [color=mycolor1, line width=1 pt,  mark=square, mark size=2.5pt,  mark options={solid, mycolor1}]
  table[row sep=crcr]{%
-20	-15.6549109952379\\
-15	-10.742692372494\\
-10	-6.02583076481947\\
-5	-1.74665506335315\\
0	1.54557778085494\\
5	3.61521145044639\\
10	4.53893837800687\\
15	4.86094758419146\\
20	5.02409329663847\\
};
\addlegendentry{\footnotesize No RIS (MC)}

\addplot [color=mycolor1, dashed, line width=1 pt,  mark=triangle, mark options={solid, mycolor1}]
  table[row sep=crcr]{%
-20	-15.7799416482315\\
-15	-10.8644138585678\\
-10	-6.12121317335858\\
-5	-1.84579045436739\\
0	1.43422142302313\\
5	3.35042840806978\\
10	4.18790147645158\\
15	4.49027816423386\\
20	4.59045191073867\\
};
\addlegendentry{\footnotesize  Corollary 2 \cite{HJY}}

\addplot [color=red, line width=1 pt,  mark=o, mark size=2.5pt, mark options={solid, red}]
  table[row sep=crcr]{%
-20	-1.61209900542249\\
-15	1.65430562191976\\
-10	3.66549196434489\\
-5	4.5357252052572\\
0	4.87366616629324\\
5	5.05157518747261\\
10	5.02575501423955\\
15	5.06648107302837\\
20	5.08244154367215\\
};
\addlegendentry{\footnotesize  With RIS (MC), $N=32$}

\addplot [color=red, dashed, line width=1 pt, mark=x, mark options={solid, red}]
  table[row sep=crcr]{%
-20	-1.70295059821393\\
-15	1.52983066300975\\
-10	3.39716787446281\\
-5	4.20576590360383\\
0	4.49632650470075\\
5	4.59240827192138\\
10	4.62323976738892\\
15	4.63303527080416\\
20	4.63613748463001\\
};
\addlegendentry{\footnotesize Corollary \ref{Cor:spec}, $N=32$}

\addplot [color=blue, line width=1 pt,mark=diamond, mark size=2.5pt, mark options={solid, blue}]
  table[row sep=crcr]{%
-20	0.458794017931727\\
-15	2.9865720407793\\
-10	4.27781087496087\\
-5	4.76986783965534\\
0	4.95370462621038\\
5	5.02443301112687\\
10	5.07287875493873\\
15	4.99807806823951\\
20	5.02730350914199\\
};
\addlegendentry{\footnotesize  With RIS (MC), $N=64$}

\addplot [color=blue, dashed, line width=1 pt,  mark=asterisk, mark options={solid, blue}]
  table[row sep=crcr]{%
-20	0.3589455665812\\
-15	2.78858156045264\\
-10	3.96376859479159\\
-5	4.41298736259552\\
0	4.56528924103724\\
5	4.61458448709943\\
10	4.63029018286877\\
15	4.63526860363242\\
20	4.63684410693146\\
};
\addlegendentry{\footnotesize  Corollary \ref{Cor:spec}, $N=64$}

\end{axis}
\end{tikzpicture}%
\caption{Average SINR under the channel model in \eqref{spec} for $L=1$, $M=32$ and $K=12$.} 
\label{Figlast}
\end{minipage}
\end{figure*}

 We plot  in Fig. \ref{Fig6} the net sum-rate in (\ref{R_sum_ray}) under Rayleigh fading using the deterministic equivalent of the SINR in Corollary \ref{th2} for MMSE-DFT CE protocol and the one in Corollary \ref{Thm2} for MMSE-DE CE protocol. These results do not depend on RIS phase shifts as discussed in Sec. III-C, indicating that RISs do not yield any reflect beamforming gains in terms of the ergodic net sum rate under Rayleigh fading. To highlight this we plot the Monte-Carlo (MC) simulated ergodic net sum-rate in (\ref{R_sum_MC})   using random RISs phase shifts as well as using the RISs phase shifts matrix $\boldsymbol{\Theta}$ that is selected from a random set of $F$ $NL\times NL$  $\boldsymbol{\Theta}_l$'s to maximize the MC simulated net sum-rate. We refer to this case as  optimized (opt.) RISs.  As expected,  the performance of the RISs-assisted system under random and optimized phase shifts is the same. However, with an increasing number of RIS elements, the RISs-assisted system still outperforms the conventional (No RISs) system  due to the array gain that the RISs yield even under Rayleigh fading.  Comparing the curves in Fig. \ref{Fig5} and Fig. \ref{Fig6}, we also observe that the net sum-rate for all considered systems is better under Rician fading due to the presence of LoS components. 

Similar to the result under Rician fading, the net sum-rate under DFT-CE protocol first increases and then starts to decrease after $N=150$ because of the increase in training overhead.   Contrary to the trend in Fig. \ref{Fig5}, the performance under MMSE-DFT CE protocol is better than that under MMSE-DE CE protocol  for the considered range of $N$.  The main reason is that unlike the case in Rician fading, the RIS phase shifts do not play a major role in improving the downlink ergodic rates  under Rayleigh fading. However, they still play an important role in the uplink to improve the CE accuracy. Optimization of the RIS phase shifts in MMSE-DFT protocol to achieve accurate estimate of $\mathbf{h}_k^{\rm ray}$ for precoding provides a significant SINR gain  over MMSE-DE protocol  for a large range of $N$. However, eventually for $N>320$ the MMSE-DE CE protocol will start outperforming the MMSE-DFT CE protocol. 


Fig. \ref{Figlast} studies the performance of RIS-assisted system against $\rho$ under the simplified channel model considered in Corollary \ref{Cor:spec} for perfect CSI. The match between the Monte-Carlo simulated average SINR and the theoretical result \eqref{special1} in Corollary \ref{Cor:spec} is good. We also observe that the RIS is only beneficial under Rayleigh fading when $\rho$ takes small to moderate values, i.e. the system is noise-limited.  For interference-limited scenarios   under Rayleigh fading (i.e. high $\rho$), the performance of RIS-assisted system approaches that of the conventional MISO system, studied in \cite{HJY}. These observations corroborate our analysis and insights from Sec. III-C.

\section{Conclusion}
In this work, we studied the  net sum-rate performance of a distributed RISs-assisted multi-user MISO  communication system under  Rician  and Rayleigh fading environments. We considered two CE protocols, namely the DFT-CE and the DE-CE protocols, and derived the MMSE  estimates of the aggregate channel under each protocol. Considering imperfect I-CSI for  precoding at the BS and given RISs phase shifts, we derived the deterministic equivalents of the SINR and achievable net sum-rate under Rician and Rayleigh fading scenarios and under each CE protocol. The RIS phase shifts were then optimized using these expressions based on S-CSI. As a benchmark, we also devised a scheme where the  phase shifts were instantaneously optimized using full I-CSI obtained using the DFT-CE protocol. Results showed that DE of the overall channel for precoding with S-CSI based design for the RISs phase shifts outperforms both DFT-CE based schemes, i.e. the one with RISs designed using S-CSI as well as the one with RISs designed using full I-CSI, owing to the significantly lower training overhead of the DE scheme.

\appendix

\subsection{Proof of Lemma \ref{L1_RIC}}\vspace{-.05in}
\label{appena4}
Recall $\hat{\mathbf{h}}^{\rm ric}_{k}= \hat{\mathbf{h}}^{\rm n}_{dk}+\bar{\mathbf{h}}_{dk}+\sum_{l=1}^L \mathbf{H}_{1l}\boldsymbol{\Theta}_l \mathbf{h}^{\rm n}_{2lk}+\sum_{l=1}^L \mathbf{H}_{1l}\boldsymbol{\Theta}_l \bar{\mathbf{h}}_{2lk} $, with mean $\mathbb{E}[\hat{\mathbf{h}}^{\rm ric}_{k}]=\bar{\mathbf{h}}_{dk}+\sum_{l=1}^L \mathbf{H}_{1l}\boldsymbol{\Theta}_l \bar{\mathbf{h}}_{2lk}$ and covariance matrix $\mathbf{C}_k^{\rm ric}=\mathbb{E}[\hat{\mathbf{h}}^{\rm ric}_k \hat{\mathbf{h}}^{ric^H}_k]-\mathbb{E}[\hat{\mathbf{h}}^{\rm ric}_k] \mathbb{E}[\hat{\mathbf{h}}^{\rm ric}_k]^H$. Noting that $\hat{\mathbf{h}}^{\rm n}_{2lk}$ and $\hat{\mathbf{h}}^{\rm n}_{dk}$ are independent, we can define the covariance matrix as \vspace{-.05in}
\begin{align}
\label{fin2}
&\mathbf{C}_k^{\rm ric}=\mathbb{E}[\hat{\mathbf{h}}^{\rm ric}_k \hat{\mathbf{h}}^{ric^H}_k]-\mathbb{E}[\hat{\mathbf{h}}^{\rm ric}_k] \mathbb{E}[\hat{\mathbf{h}}^{\rm ric}_k]^H=\mathbb{E}[\hat{\mathbf{h}}^{\rm n}_{dk}\hat{\mathbf{h}}_{dk}^{n^H}]+\mathbb{E}[\sum_{l=1}^L \sum_{l'=1}^L \mathbf{H}_{1l}\boldsymbol{\Theta}_l \hat{\mathbf{h}}^{\rm n}_{2lk} \hat{\mathbf{h}}^{n^H}_{2l'k}\boldsymbol{\Theta}_{l'}^{H}\mathbf{H}_{1l'}^H].
\end{align} \normalsize
We compute $\mathbb{E}[\hat{\mathbf{h}}^{\rm n}_{dk}\hat{\mathbf{h}}_{dk}^{n^H}]=\frac{\beta_{dk}^{n^2} \mathbb{E}[(\tilde{\mathbf{r}}_{0k}^{tr}-\bar{\mathbf{h}}_{dk})(\tilde{\mathbf{r}}_{0k}^{tr}-\bar{\mathbf{h}}_{dk})^H]}{(\beta_{dk}^{\rm n}+\frac{1}{S \rho_p \tau_S})^2}=\frac{\beta_{dk}^{n^2} (\beta_{dk}^{\rm n}+\frac{1}{S\rho_{p} \tau_S})}{(\beta_{dk}^{\rm n}+\frac{1}{S\rho_{p}\tau_S})^2} \mathbf{I}_M=\frac{\beta_{dk}^{n^2}}{\beta_{dk}^{\rm n}+\frac{1}{S\rho_{p}\tau_S}}\mathbf{I}_M$ using the definitions of $\hat{\mathbf{h}}^{\rm n}_{dk}$ and $\tilde{\mathbf{r}}_{0k}^{tr}$ in Lemma 1. Next compute $\mathbb{E}[\sum_{l=1}^L \sum_{l'=1}^L \mathbf{H}_{1l}\boldsymbol{\Theta}_l \hat{\mathbf{h}}^{\rm n}_{2lk} \hat{\mathbf{h}}^{n^H}_{2l'k}\boldsymbol{\Theta}_{l'}^{H}\mathbf{H}_{1l'}^H]$ by using the expression of $\hat{\mathbf{h}}^{\rm n}_{2lk}$ from \eqref{h_d_est_RIC} and noting the independence between $\hat{\mathbf{h}}^{\rm n}_{2lk}$ for $l\neq l'$ as \vspace{-.05in}
\begin{align}
&\mathbb{E}[\sum_{l=1}^L \sum_{l'=1}^L \mathbf{H}_{1l}\boldsymbol{\Theta}_l \hat{\mathbf{h}}^{\rm n}_{2lk} \hat{\mathbf{h}}^{n^H}_{2l'k}\boldsymbol{\Theta}_{l'}^{H}\mathbf{H}_{1l'}^H]=\sum_{l=1}^L \mathbf{H}_{1l}\boldsymbol{\Theta}_l  \mathbb{E}[\hat{\mathbf{h}}^{\rm n}_{2lk} \hat{\mathbf{h}}^{n^H}_{2lk}]\boldsymbol{\Theta}_{l}^{H}\mathbf{H}_{1l}^H, \\
\label{fin1}
&\overset{(a)}{=}  \sum_{l=1}^L \left(\frac{\beta_{2lk}^{n^2}}{\beta_{2lk}^{\rm n}+\frac{1}{S\rho_{p}\tau_S} M \beta_{1l}}\right) \mathbf{H}_{1l}\boldsymbol{\Theta}_l \boldsymbol{\Theta}_{l}^{H}\mathbf{H}_{1l}^H\overset{(b)}{=}  \sum_{l=1}^L \left(\frac{\beta_{2lk}^{n^2}}{\beta_{2lk}^{\rm n}+\frac{1}{S\rho_{p}\tau_S} M \beta_{1l}}\right) \mathbf{H}_{1l}\mathbf{H}_{1l}^H
\end{align}
where (a) is obtained by computing $\mathbb{E}[\hat{\mathbf{h}}^{\rm n}_{2lk}\hat{\mathbf{h}}_{2lk}^{n^H}]=\frac{\beta_{2lk}^{n^2}}{(\beta_{2lk}^{\rm n}+\frac{1}{S \rho_p \tau_S M \beta_{1l}})^2}\mathbb{E}[(\tilde{\mathbf{r}}_{lk}^{tr}-\bar{\mathbf{h}}_{2lk})(\tilde{\mathbf{r}}_{lk}^{tr}-\bar{\mathbf{h}}_{2lk})^H]=\frac{\beta_{2lk}^{n^2}}{\beta_{2lk}^{\rm n}+\frac{1}{S\rho_{p}\tau_S M \beta_{1l}}}\mathbf{I}_N$.  Also (b) is obtained by noting that $\boldsymbol{\Theta}_l \boldsymbol{\Theta}_{l}^{H}=\textbf{I}_N$. Putting \eqref{fin1} and $\mathbb{E}[\hat{\mathbf{h}}^{\rm n}_{dk}\hat{\mathbf{h}}_{dk}^{n^H}]=\frac{\beta_{dk}^{n^2}}{\beta_{dk}^{\rm n}+\frac{1}{S\rho_{p} \tau_S}}\mathbf{I}_M$ together in \eqref{fin2} yields the expression of $\mathbf{C}_k^{\rm ric}$. The proof is  completed by  defining $\hat{\mathbf{h}}^{\rm ric}_{k}$ as a complex Gaussian vector with mean $\bar{\mathbf{h}}_{dk}+\sum_{l=1}^L \mathbf{H}_{1l}\boldsymbol{\Theta}_l \bar{\mathbf{h}}_{2lk}$ and covariance $\mathbf{C}_k^{\rm ric}$.

\subsection{Proof of Theorem \ref{thm1_ric}}
\label{appenb_ric}
The proof starts by dividing the numerator and denominator of (\ref{SINR_MRT}) by $M$ and working separately on the four terms: 1)  scaled signal power $p_k|\mathbb{E}[\frac{1}{M}\mathbf{h}_{k}^{\rm ric^H}\hat{\mathbf{h}}^{\rm ric}_{k}]|^2$, 2)  scaled interference power $\frac{1}{M}\sum_{f\neq k}\frac{p_f}{M}\mathbb{E}[|\mathbf{h}_{k}^{\rm ric^H}\hat{\mathbf{h}}_{f}^{\rm ric}|^2]$,  3) power normalization  $\frac{1}{M^2}\Psi$ and 4) variance term $\frac{p_{k}}{M^2} \mathbb{V}\text{ar}[\mathbf{h}_{k}^{\rm ric^H} \hat{\mathbf{h}}_{k}^{\rm ric}]$. 

\subsubsection{Deterministic equivalent of $\frac{1}{M}\mathbf{h}_{k}^{\rm ric^H}\hat{\mathbf{h}}^{\rm ric}_{k}$ } 

Using the definition of the RIS-assisted channel under Rician fading in \eqref{eq_ch1_RIC}, we write $\frac{1}{M}\mathbf{h}_{k}^{ric^H}\hat{\mathbf{h}}_{k}^{\rm ric}= \frac{1}{M}(\mathbf{h}_{dk}^{\rm n}+\bar{\mathbf{h}}_{dk}+\sum_{l=1}^{L}\mathbf{H}_{1l}\boldsymbol{\Theta}_l \mathbf{h}_{2lk}^{\rm n}+\sum_{l=1}^{L}\mathbf{H}_{1l}\boldsymbol{\Theta}_l \bar{\mathbf{h}}_{2lk})^H(\hat{\mathbf{h}}_{dk}^{\rm n}+\bar{\mathbf{h}}_{dk}+\sum_{l=1}^{L}\mathbf{H}_{1l}\boldsymbol{\Theta}_l \hat{\mathbf{h}}_{2lk}^{\rm n}+\sum_{l=1}^{L}\mathbf{H}_{1l}\boldsymbol{\Theta}_l \bar{\mathbf{h}}_{2lk})$. Noting that $\hat{\mathbf{h}}^{\rm n}_{dk}$ and $\mathbf{h}^{\rm n}_{2lk}$  are independent vectors,  ${\mathbf{h}}^{\rm n}_{dk}$ and $\hat{\mathbf{h}}^{\rm n}_{2lk}$ are  independent vectors, and using the result that quadratic forms with one deterministic vector and one random zero mean vector converge to $0$, as well as applying \cite[Lemma 4 (iii)]{HJY}, we get
\begin{align}
\label{steppp}
&\frac{1}{M}\mathbf{h}_{k}^{ric^H}\hat{\mathbf{h}}^{\rm ric}_{k}- \frac{1}{M} \left(\text{tr }\mathbf{D}_{k}+\mathbf{h}_{dk}^{n^H}\hat{\mathbf{h}}_{dk}^{\rm n}+\sum_{l=1}^{L}\sum_{l'=1}^{L} {\mathbf{h}}_{2lk}^{n^H} \boldsymbol{\Theta}_l^H \mathbf{H}_{1l}^H \mathbf{H}_{1l'}\boldsymbol{\Theta}_{l'} \hat{\mathbf{h}}_{2l'k}^{\rm n}\right)\xrightarrow[M,N,K\rightarrow \infty]{a.s.} 0.
\end{align}
where $\mathbf{D}_{k}=\bar{\mathbf{h}}_{dk}\bar{\mathbf{h}}^{H}_{dk}+\bar{\mathbf{h}}_{dk}\sum_{l=1}^{L}\bar{\mathbf{h}}_{2lk}^{{H}} \boldsymbol{\Theta}_l^H \mathbf{H}_{1l}^H+\sum_{l=1}^{L}\mathbf{H}_{1l} \boldsymbol{\Theta}_l \bar{\mathbf{h}}_{2lk} \bar{\mathbf{h}}_{dk}^{H}\\+\sum_{l=1}^{L}\sum_{l'=1}^{L}\mathbf{H}_{1l} \boldsymbol{\Theta}_l \bar{\mathbf{h}}_{2lk}\bar{\mathbf{h}}_{2l'k}^{{H}} \boldsymbol{\Theta}_{l'}^H \mathbf{H}_{1l'}^H$.

We work with the two random terms in (\ref{steppp}) separately. Using (\ref{h_d_est_RIC}), we obtain
\begin{align}
\label{NUM_RAY_F}
\frac{1}{M}\mathbf{h}_{dk}^{n^H}\hat{\mathbf{h}}^{\rm n}_{dk}=\frac{1}{M}\frac{\beta_{dk}^{\rm n}}{\beta_{dk}^{\rm n}+\frac{1}{S\rho_{p}\tau_S}} \mathbf{h}_{dk}^{n^H}(\mathbf{h}_{dk}^{\rm n}+\frac{1}{S} (\mathbf{v}_1^{tr} \otimes \mathbf{I}_{M})^H \mathbf{{n}}^{tr}_{k})
\end{align}

Since $\frac{1}{S} (\mathbf{v}_1^{tr} \otimes \mathbf{I}_{M})^H \mathbf{{n}}^{tr}_{k}$ and $ \mathbf{h}_{dk}^{n^H}$ are independent, we apply \cite[Lemma 4 (iii)]{HJY} to get $\frac{1}{M}\mathbf{h}_{dk}^{n^H}\hat{\mathbf{h}}_{dk}^{\rm n}-\frac{1}{M}\frac{\beta_{dk}^{\rm n}}{\beta_{dk}^{\rm n}+\frac{1}{S\rho_{p} \tau_S}}\mathbf{h}_{dk}^{n^H}\mathbf{h}_{dk}^{\rm n}\xrightarrow[M,N,K\rightarrow \infty]{a.s.} 0$. Given $\mathbf{h}_{dk}\sim \mathcal{CN}(\mathbf{0},\beta_{dk}^{\rm n}\mathbf{I}_M)$, we apply \cite[Lemma 4 (ii)]{HJY} to obtain,\vspace{-.15in}
\begin{align}
\label{11}
\frac{1}{M}\mathbf{h}_{dk}^{n^H}\hat{\mathbf{h}}_{dk}^{\rm n}-\frac{1}{M}\frac{\beta_{dk}^{n^2}M}{\beta_{dk}^{\rm n}+\frac{1}{S\rho_{p} \tau_S}}\xrightarrow[M,N,K\rightarrow \infty]{a.s.} 0.
\end{align}

Next we work on the second term  in (\ref{steppp}) using $\hat{\mathbf{h}}_{2lk}^{\rm n}$ from \eqref{h_d_est_RIC} to write $
\sum_{l=1}^{L}\sum_{l'=1}^{L} {\mathbf{h}}_{2lk}^{n^H} \boldsymbol{\Theta}_l^H \mathbf{H}_{1l}^H \\
\mathbf{H}_{1l'}\boldsymbol{\Theta}_{l'} \hat{\mathbf{h}}_{2l'k}^{\rm n}=\sum_{l=1}^{L}\sum_{l'=1}^{L} {\mathbf{h}}_{2lk}^{n^H} \boldsymbol{\Theta}_l^H \mathbf{H}_{1l}^H \mathbf{H}_{1l'}\boldsymbol{\Theta}_{l'} \frac{\beta_{2l'k}^{\rm n}}{\beta_{2l'k}^{\rm n}+\frac{1}{S\rho_{p}\tau_S M \beta_{1l'}}} \left(\mathbf{h}_{2l'k}^{\rm n}+\frac{\bar{\mathbf{H}}_{1l'}^H(\mathbf{V}_{l'}^{tr} \otimes \mathbf{I}_{M})^H \mathbf{{n}}^{tr}_{k}}{SM\beta_{1l'}}\right)$. Noting that $\mathbf{h}_{2lk}^{\rm n}$ is independent of $\frac{1}{SM\beta_{1l'}}\bar{\mathbf{H}}_{1l'}^H(\mathbf{V}_{l'}^{tr} \otimes \mathbf{I}_{M})^H \mathbf{{n}}^{tr}_{k}$, and independent of $\mathbf{h}_{2l'k}^{\rm n}$ for $l \neq l'$, we apply  \cite[Lemma 4 (iii)]{HJY} to obtain \small
\begin{align}
&\frac{1}{M}\sum_{l=1}^{L}\sum_{l'=1}^{L} {\mathbf{h}}_{2lk}^{n^H} \boldsymbol{\Theta}_l^H \mathbf{H}_{1l}^H \mathbf{H}_{1l'}\boldsymbol{\Theta}_{l'} \hat{\mathbf{h}}_{2l'k}^{\rm n}-\frac{1}{M}\sum_{l=1}^{L} {\mathbf{h}}_{2lk}^{n^H} \boldsymbol{\Theta}_l^H \mathbf{H}_{1l}^H \mathbf{H}_{1l}\boldsymbol{\Theta}_{l}  \left(\frac{\beta_{2lk}^{\rm n}}{\beta_{2lk}^{\rm n}+\frac{1}{S\rho_{p}\tau_S M \beta_{1l}}} \right) \mathbf{h}_{2lk}^{\rm n} \xrightarrow[M,N,K\rightarrow \infty]{a.s.} 0.
\end{align} \normalsize
Applying \cite[Lemma 4 (ii)]{HJY} on the quadratic term in $\mathbf{h}_{2lk}\sim \mathcal{CN}(\mathbf{0},\beta_{2lk}^{\rm n}\mathbf{I}_N)$ and noting that $\text{tr}(\boldsymbol{\Theta}_l^H \mathbf{H}_{1l}^H \mathbf{H}_{1l}\boldsymbol{\Theta}_{l})=\text{tr}(\mathbf{H}_{1l}\boldsymbol{\Theta}_{l}\boldsymbol{\Theta}_l^H \mathbf{H}_{1l}^H)=\text{tr}(\mathbf{H}_{1l} \mathbf{H}_{1l}^H)$ we obtain \small
\begin{align}
\label{33}
&\frac{1}{M}\sum_{l=1}^{L} {\mathbf{h}}_{2lk}^{n^H} \boldsymbol{\Theta}_l^H \mathbf{H}_{1l}^H \mathbf{H}_{1l}\boldsymbol{\Theta}_{l}\left(\frac{\beta_{2lk}^{\rm n}}{\beta_{2lk}^{\rm n}+\frac{1}{S\rho_{p}\tau_S M \beta_{1l}}} \right) \mathbf{h}_{2lk}^{\rm n}-\frac{1}{M}\sum_{l=1}^{L}\text{tr}(\mathbf{H}_{1l}\mathbf{H}_{1l}^H) \left(\frac{\beta_{2lk}^{n^2}}{\beta_{2lk}^{\rm n}+\frac{1}{S\rho_{p}\tau_S M \beta_{1l}}} \right)\xrightarrow[M,N,K\rightarrow \infty]{a.s.} 0.
\end{align} \normalsize

Therefore putting \eqref{11} and \eqref{33} together in \eqref{steppp} we obtain \small
\begin{align*}
&\frac{1}{M}\mathbf{h}_{k}^{ric^H}\hat{\mathbf{h}}_{k}^{\rm ric}-\frac{1}{M}\text{tr}\left(\mathbf{D}_{k}+\frac{\beta_{dk}^{n^2}}{\beta_{dk}^{\rm n}+\frac{1}{S\rho_p \tau_S}}\boldsymbol{I}_{M}+\sum_{l=1}^{L}(\mathbf{H}_{1l}\mathbf{H}_{1l}^H) \left(\frac{\beta_{2lk}^{n^2}}{\beta_{2lk}^{\rm n}+\frac{1}{S\rho_{p}\tau_S M \beta_{1l}}} \right)\right)\xrightarrow[M,N,K\rightarrow \infty]{a.s} 0 \nonumber.
\end{align*}\normalsize

\subsubsection{Deterministic equivalent of $\frac{1}{M}\sum_{f\neq k}\frac{p_{f}}{M}|\mathbf{h}^{\rm ric^H}_{k}\hat{\mathbf{h}}^{\rm ric}_{f}|^2 $ }
\label{Det_1}
Note that $\frac{1}{M}\sum_{f\neq k}\frac{p_{f}}{M}|\mathbf{h}_{k}^{ric^H}\hat{\mathbf{h}}_{f}^{\rm ric}|^2=\frac{1}{M^2}\mathbf{h}_{k}^{ric^H}\hat{\mathbf{H}}^{ric^H}_{[k]}\mathbf{P}_{[k]}\hat{\mathbf{H}}^{\rm ric}_{[k]}\mathbf{h}_{k}^{\rm ric}$, where $\hat{\mathbf{H}}^{\rm ric}_{[k]}=[\hat{\mathbf{h}}_1^{\rm ric}, \dots, \hat{\mathbf{h}}_{k-1}^{\rm ric}, \hat{\mathbf{h}}_{k+1}^{\rm ric}, \dots, \hat{\mathbf{h}}_K^{\rm ric}]^H$. Note that \\ $\hat{\mathbf{H}}^{ric^H}_{[k]}\mathbf{P}_{[k]}\hat{\mathbf{H}}^{\rm ric}_{[k]}$ is independent of $\mathbf{h}_k^{\rm ric}$. Moreover $\mathbf{h}_k^{\rm ric}\sim \mathcal{CN}(\boldsymbol{\mu}_k^{\rm ric}, \mathbf{A}^{\rm ric}_k)$ where $\boldsymbol{\mu}_k^{\rm ric}=\bar{\mathbf{h}}_{dk}+\sum_{l=1}^{L}\mathbf{H}_{1l}\boldsymbol{\Theta}_l\bar{\mathbf{h}}_{2lk}$ and $\mathbf{A}^{\rm ric}_k$ is defined in \eqref{eq_ch1_RIC22}. Using these observations and defining $\mathbf{D}_{k}=\boldsymbol{\mu}_k^{\rm ric}\boldsymbol{\mu}_k^{\rm ric^H}$, we apply \cite[Lemma 4 (ii)]{HJY} to obtain
\begin{align}
\frac{1}{M^2}\mathbf{h}_{k}^{ric^H}\hat{\mathbf{H}}^{ric^H}_{[k]}\mathbf{P}_{[k]}\hat{\mathbf{H}}^{\rm ric}_{[k]}\mathbf{h}_{k}^{\rm ric}-\frac{1}{M^2}\text{tr}((\mathbf{D}_{k}+\mathbf{A}^{\rm ric}_{k})\hat{\mathbf{H}}^{ric^H}_{[k]}\textbf{P}_{[k]}\hat{\mathbf{H}}^{\rm ric}_{[k]})\xrightarrow[M,N,K\rightarrow \infty]{a.s} 0.
\end{align}

Note that $\text{tr}((\mathbf{D}_{k}+\mathbf{A}^{\rm ric}_{k})\hat{\mathbf{H}}^{ric^H}_{[k]}\mathbf{P}_{[k]}\hat{\mathbf{H}}^{\rm ric}_{[k]})=\sum_{f\neq k}p_{f}\hat{\mathbf{h}}_{f}^{ric^H}(\mathbf{D}_{k}+\mathbf{A}^{\rm ric}_{k})\hat{\mathbf{h}}_{f}^{\rm ric}$. Applying \cite[Lemma 4 (ii)]{HJY} on the quadratic form in $\hat{\mathbf{h}}_{f}^{\rm ric}\sim \mathcal{CN}(\boldsymbol{\mu}_f^{\rm ric}, \mathbf{C}^{\rm ric}_f)$  \eqref{est_corr_RIC}, yields
\begin{align}
\frac{1}{M}\sum_{f\neq k}\frac{p_{f}}{M}\hat{\mathbf{h}}_{f}^{ric^H}(\mathbf{D}_{k}+\mathbf{A}_{k}^{\rm ric})\hat{\mathbf{h}}_{f}^{\rm ric}-\frac{1}{M}\sum_{f\neq k}\frac{p_{f}}{M}\text{tr}((\mathbf{D}_{f}+\mathbf{C}_{f}^{\rm ric})(\mathbf{D}_{k}+\mathbf{A}_{k}^{\rm ric}))\xrightarrow[M,N,K\rightarrow \infty]{a.s} 0
\end{align}
where  $\mathbf{C}_{f}^{\rm ric}$ is defined in Lemma 2. Therefore we obtain \vspace{-.1in}
\begin{align}
\frac{1}{M}\sum_{f\neq k}\frac{p_{f}}{M}|\mathbf{h}_{k}^{ric^H}\hat{\mathbf{h}}_{f}^{\rm ric}|^2-\frac{1}{M}\sum_{f\neq k}\frac{p_{f}}{M}\text{tr}((\mathbf{D}_{f}+\mathbf{C}_{f}^{\rm ric})(\mathbf{D}_{k}+\mathbf{A}_{k}^{\rm ric}))\xrightarrow[M,N,K\rightarrow \infty]{a.s} 0
\end{align}

\subsubsection{Deterministic equivalent of $\frac{1}{M^2}\Psi=\frac{1}{M^2} \text{tr}(\mathbf{P}\hat{\mathbf{H}^{\rm ric}}\hat{\mathbf{H}}^{\rm ric^H})$ }
Note that  $\frac{1}{M^2}\Psi=\\ \frac{1}{M}\sum_{k=1}^{K}\frac{p_{k}}{M}\hat{\mathbf{h}}_{k}^{ric^H}\hat{\mathbf{h}}_{k}^{\rm ric}$. Applying  \cite[Lemma 4 (ii)]{HJY} on the quadratic form in $\hat{\mathbf{h}}_{k}^{\rm ric}\sim \mathcal{CN}(\boldsymbol{\mu}_k^{\rm ric}, \mathbf{C}^{\rm ric}_k)$, we obtain $\frac{1}{M}\sum_{k=1}^{K}\frac{p_{k}}{M}\hat{\mathbf{h}}_{k}^{ric^H}\hat{\mathbf{h}}_{k}^{\rm ric}-\frac{1}{M}\sum_{k=1}^{K}\frac{p_{k}}{M}\text{tr}(\mathbf{D}_{k}+\mathbf{C}_{k}^{\rm ric})\xrightarrow[M,N,K\rightarrow \infty]{a.s} 0.$

\subsubsection{Deterministic equivalent of $\frac{p_{k}}{M^2} \mathbb{V}\text{ar}[\mathbf{h}^{\rm ric^{H}}_{k} \hat{\mathbf{h}}^{\rm ric}_{k}]$}

 Note that $\frac{p_{k}}{M^2} \mathbb{V}\text{ar}[\mathbf{h}_{k}^{ric^H} \hat{\mathbf{h}}_{k}^{\rm ric}]=\mathbb{V}\text{ar}(x+y+\bar{x}+\bar{y})$, where $x=\frac{p_{k}}{M^2}\mathbf{h}_{k}^{ric^H} \hat{\mathbf{h}}_{dk}^{\rm n} $, $\bar{x}=\frac{p_{k}}{M^2}\mathbf{h}_{k}^{ric^H} \bar{\mathbf{h}}_{dk} $,  $y=\frac{p_{k}}{M^2}\mathbf{h}_{k}^{ric^H} \sum_{l=1}^L \mathbf{H}_{1l}\boldsymbol{\Theta}_l \hat{\mathbf{h}}_{2lk}^{\rm n}$ and $\bar{y}=\frac{p_{k}}{M^2}\mathbf{h}_{k}^{ric^H} \sum_{l=1}^L \mathbf{H}_{1l}\boldsymbol{\Theta}_l \bar{\mathbf{h}}_{2lk}$. We know that $\mathbb{V}\text{ar}(x+y+\bar{x}+\bar{y})\leq \mathbb{V}\text{ar}(x)+ \mathbb{V}\text{ar}(y)+\mathbb{V}\text{ar}(\bar{x})+ \mathbb{V}\text{ar}(\bar{y})$. The variance of $x$ is bounded by $\mathbb{E}[|x|^2]$ which converges to $0$. To see this note that $\frac{p_{k}}{M}\mathbf{h}_{k}^{ric^H} \hat{\mathbf{h}}_{dk}^{\rm n} -\frac{p_{k}}{M} \frac{\beta_{dk}^{n^2} }{\beta_{dk}^{\rm n}+\frac{1}{S\rho_{p}\tau_S}}\text{tr}(\mathbf{I}_M)\xrightarrow[M,K\rightarrow \infty]{a.s} 0$ by using the same steps as done to get (40). Then  $\mathbb{E}[|x|^2]=\frac{p_{k}^2 M^2}{M^4} \left( \frac{\beta_{dk}^{n^2}}{\beta_{dk}^{\rm n}+\frac{1}{S\rho_{p}\tau_S}} \right)^2 +o(1)$, and $\mathbb{E}[|x|^2]\xrightarrow[M,K\rightarrow \infty]{a.s} 0$. The variance of $y$, $\bar{x}$ and $\bar{y}$ can also be proved to converge to zero similarly. Therefore $\frac{p_{k}}{M^2} \mathbb{V}\text{ar}[\mathbf{h}_{k}^{ric^H} \hat{\mathbf{h}}_{k}^{\rm ric}]\xrightarrow[M,K\rightarrow \infty]{a.s} 0$.  

Combining the results of these subsections completes the proof of Theorem 1.
\vspace{-.1in}
\bibliographystyle{IEEEtran}
\bibliography{bibl}

\end{document}